\documentclass[]{article}
\usepackage{graphicx}
\usepackage{hyperref}
\usepackage{multirow}
\usepackage{amsmath,amssymb,amsfonts}
\usepackage{amsthm}
\usepackage{mathrsfs}
\usepackage[title]{appendix}
\usepackage{xcolor}
\usepackage{textcomp}
\usepackage{booktabs}
\usepackage{algorithm}
\usepackage{algorithmicx}
\usepackage{algpseudocode}
\usepackage{tablefootnote}
\usepackage{listings}
\usepackage{makecell}
\usepackage{float}
\usepackage{placeins}
\usepackage{authblk}
\usepackage[font=footnotesize]{caption}
\usepackage[a4paper,top=1in,bottom=1in,left=1.0in,right=1.0in]{geometry}
\usepackage[numbers]{natbib}  
\usepackage{orcidlink}

\begin{document}
\title{\huge\bfseries The Quantum Ensemble Variational Optimization Algorithm: Applications to Molecular Inverse Design}

\author[*,1,2,3]{Francesco Calcagno \orcidlink{0000-0002-0986-4425}}  
\author[3]{Delmar G. A. Cabral \orcidlink{0009-0001-1195-5529}}
\author[*,1,2]{Ivan Rivalta \orcidlink{0000-0002-1208-602X}} 
\author[*,3,4]{Victor S. Batista \orcidlink{0000-0002-3262-1237}}
\affil[1]{Department of Industrial Chemistry, Alma Mater Studiorum University of Bologna, via Piero Gobetti 85, Bologna, 40129, Italy} 
\affil[2]{Center for Chemical Catalysis - C3, Alma Mater Studiorum University of Bologna, via Piero Gobetti 85, 40129 Bologna, Italy}
\affil[3]{Department of Chemistry, Yale University, 225 Prospect St, New Haven, 06520, CT, U.S.A.}
\affil[4]{Yale Quantum Institute, Yale University, 17 Hillhouse Ave, New Haven, 06511, CT, U.S.A.}

\date{*Corresponding authors. E-mails: \href{mailto:francesco.calcagno@unibo.it}{francesco.calcagno@unibo.it}; \href{mailto:i.rivalta@unibo.it}{i.rivalta@unibo.it}; \href{mailto:victor.batista@yale.edu}{victor.batista@yale.edu}} 
\maketitle

\begin{abstract}
Designing molecules with optimized properties remains a fundamental challenge due to the intricate relationship between molecular structure and properties. Traditional computational approaches that address the combinatorial number of possible molecular designs become unfeasible as the molecular size increases, suffering from the so-called `curse of dimensionality' problem. Recent advances in quantum computing hardware present new opportunities to address this problem. Here, we introduce the Quantum Ensemble Variational Optimization (QEVO) method for near-term and early fault-tolerant quantum computing platforms. QEVO efficiently maps molecular structures onto an orthonormal basis of Pauli strings and samples from a superposition state generated by a variational ansatz. The ansatz is iteratively optimized to identify molecular candidates with the desired property. Our numerical simulations demonstrate the potential of QEVO in designing drug-like molecules with anticancer properties, employing a shallow quantum circuit that requires only a modest number of qubits. We envision that QEVO could be applied to a wide range of complex problems, offering practical solutions to problems with combinatorial complexity.
\end{abstract}

\newpage
The computational design of molecules with tailored properties remains a central and long-standing challenge in chemistry \cite{freeze_ID, aspuru-perspective, materialsID, ID_Holy_Grail, wang2006designing, xiao2008inverse, de2012inverse, tio2_VSB}. Unlike traditional approaches that search molecular databases for candidates with desired properties, molecular {\em inverse design} aims to directly generate molecular structures that meet suitable properties for specific application \cite{freeze_ID}. This paradigm underlies a broad spectrum of research efforts, spanning from drug discovery to advanced materials design.

The difficulty of inverse design lies in the inherently complex and often non-intuitive relationships between molecular structure and properties. As a result, many current methods rely on `brute-force' strategies, including exhaustive searches and high-throughput screening \cite{freedom_design}. However, the combinatorial explosion of possible molecular frameworks—the entire chemical space \cite{CS}—makes such approaches ineffective in most practical settings.

To date, a unique approach for molecular design remains elusive. As a result, numerous methodologies have been proposed, including physics-based methods \cite{tio2_VSB, ni_VSB,reiher_gdmc, reiher_gdmc2}, and computational high-throughput screening strategies. Recently, machine learning (ML) methods have emerged as valuable tools to accelerate the generation of molecules with predefined properties \cite{butler2018, sanchezlengeling2018, schwalbekoda2020, bilodeau2022, anstine2023, jcc_rl,calcagno2025quantumchemistrydrivenmolecular}. However, ML methods typically require large training datasets and do not guarantee a thorough exploration of the chemical space \cite{calcagno2025quantumchemistrydrivenmolecular}. 

Quantum computers (QCs) present new opportunities for developing computational methods that surpass the capabilities of classical systems \cite{quantum_review}. The advantage of QCs—referred to as quantum utility—extends beyond potential computational speedups, including the development of fundamentally new approaches as demonstrated by prominent quantum algorithms for search \cite{grover}, factorization \cite{shor}, and optimization \cite{VQE, qaoa, kyaw2023boosting, SQD}. 

Variational Quantum Algorithms (VQAs) employ a hybrid quantum-classical framework to tackle optimization problems and have attracted considerable attention due to their practical implementation on noisy intermediate-scale quantum (NISQ) devices and early fault-tolerant quantum computers~\cite{VQAs}. In this approach, a parameterized quantum state
\begin{equation}
    |\psi(\vec{\theta})\rangle = U(\vec{\theta})|0\rangle
\end{equation}
is prepared by applying a sequence of unitary operations \( U(\vec{\theta}) \), governed by a set of variational parameters \( \vec{\theta} \), to the initial vacuum state \( |0\rangle \). Measurements on the resulting state are processed by a classical optimizer, which iteratively updates \( \vec{\theta} \) to minimize an objective function, typically the expectation value of a Hamiltonian. This iterative procedure continues until convergence to an optimal solution is achieved~\cite{VQAs}. The method can also be extended to quantum search tasks, as in the Variational Quantum Search (VQS) approach~\cite{zhan2023}, by encoding the target solution as the ground state of an effective Hamiltonian.

However, many relevant optimization problems—such as predicting molecular solubility, toxicity, or synthetic accessibility—do not naturally lend themselves to expectation value formulations over quantum states. To address these cases, it is necessary to generalize the VQA framework to accommodate broader classes of objective functions beyond expectation values.

Here, we introduce the so-called {\em Quantum Ensemble Variational Optimization} (QEVO) algorithm, which extends VQAs to cases where the property of interest cannot be computed as an expectation value of a superposition state. QEVO optimizes the variational parameters of a quantum circuit to optimize the {\em ensemble average} of the property of interest, after sampling an ensemble of molecules from a distribution defined by the generated superposition state. We demonstrate the capabilities of QEVO by applying it to the generation of molecules optimized for specific properties, including molecular solubility, drug-likeness, and binding affinity to a protein target relevant to cancer treatment.

\section{The QEVO Algorithm}
QEVO maps molecular structures onto an orthonormal basis, $\mathcal{M}$, of binary strings $|m_i \rangle$, as described in Sec. \ref{methods}.
Each string represents a specific molecule with well-defined property $p_i$. The strings are sampled by measuring the superposition state $|\psi(\vec{\theta}) \rangle$ in the computational basis. The ensemble of strings is described by the mixed-state density matrix, $\hat{\rho}(\vec{\theta}) = \sum_i \omega_i(\vec{\theta}) | m_i \rangle \langle m_i |$, where the probability weight $\omega_i(\vec{\theta})=N_i/N$ is defined by the number of times $N_i$ the string $| m_i \rangle$ has been sampled out of the total number $N$ of sampled strings. Sampling involves measuring the variational quantum state $| M(\vec{\theta}) \rangle=U(\vec{\theta}) |0 \rangle$ in the computational basis (Fig. \ref{panel1}A). 

We use a real amplitude ansatz (Fig. \ref{panel1}, D and F), so the resulting variational state $| M(\vec{\theta}) \rangle$ is an alchemical superposition of possible molecules $| m_i  \rangle$:
\begin{equation}
    |{M(\vec{\theta})} \rangle = \sum_i \omega_i(\vec{\theta}) | m_i \rangle, \quad \mathrm{where} \quad | m_i \rangle \in \mathcal{M},
    \label{eq:M}
\end{equation}
\noindent
where $\omega_i(\vec{\theta})$ determines the probability of sampling molecule $i$ encoded by $| m_i \rangle$. 
Therefore, QEVO does not optimize the variational parameters with respect to an expectation value as in conventional VQAs \cite{VQAs,zhan2023}. Instead, QEVO optimizes $\vec{\theta}$ with respect to the loss function, 
\begin{equation}
\mathcal{L}(\vec{\theta}) = |p_M(\vec{\theta}) - p_0| + \epsilon(\vec{\theta}),
\label{eq:loss}
\end{equation}
defined by the {\em ensemble average}, 
\begin{equation}
p_M(\vec{\theta}) = \sum_i  \omega_i(\vec{\theta}) p_i.
\label{eq:density}
\end{equation}
Here, $p_i$ is the value of the property of interest for molecule $|m_i \rangle$, while $p_0$ is the target value. The ensemble average $p_M$ is computed at an arbitrary level of theory after decoding the sampled strings $| m_i \rangle$ into a molecular structure. The regularization term $\epsilon(\vec{\theta})=\lambda \cdot \omega_i(\vec{\theta})^2$ enhances the purity of $\hat{\rho}(\vec{\theta})$, as weighted by the hyperparameter $\lambda$. According to this approach, QEVO generalizes the quantum variational optimization scheme to problems where the expectation value of the property of interest cannot be computed directly from $| {M(\vec{\theta} \rangle)}$. Upon convergence, molecules $|m_i\rangle$ with the desired property can be obtained by measurements of $ | M(\vec{\theta}) \rangle$ in the computational basis, effectively bridging variational quantum optimization with generative molecular design. 

\section{Results and discussion}

\subsection{Inverse design with QEVO}
We evaluated QEVO in the context of molecular design targeting specific properties, including molecular solubility, predicted octanol-water partition coefficient (plogP), and drug-likeness (see Sec.~\ref{methods} for computational details). 

\begin{figure}[!htbp]
    \centering
    \includegraphics[width=.83\textwidth]{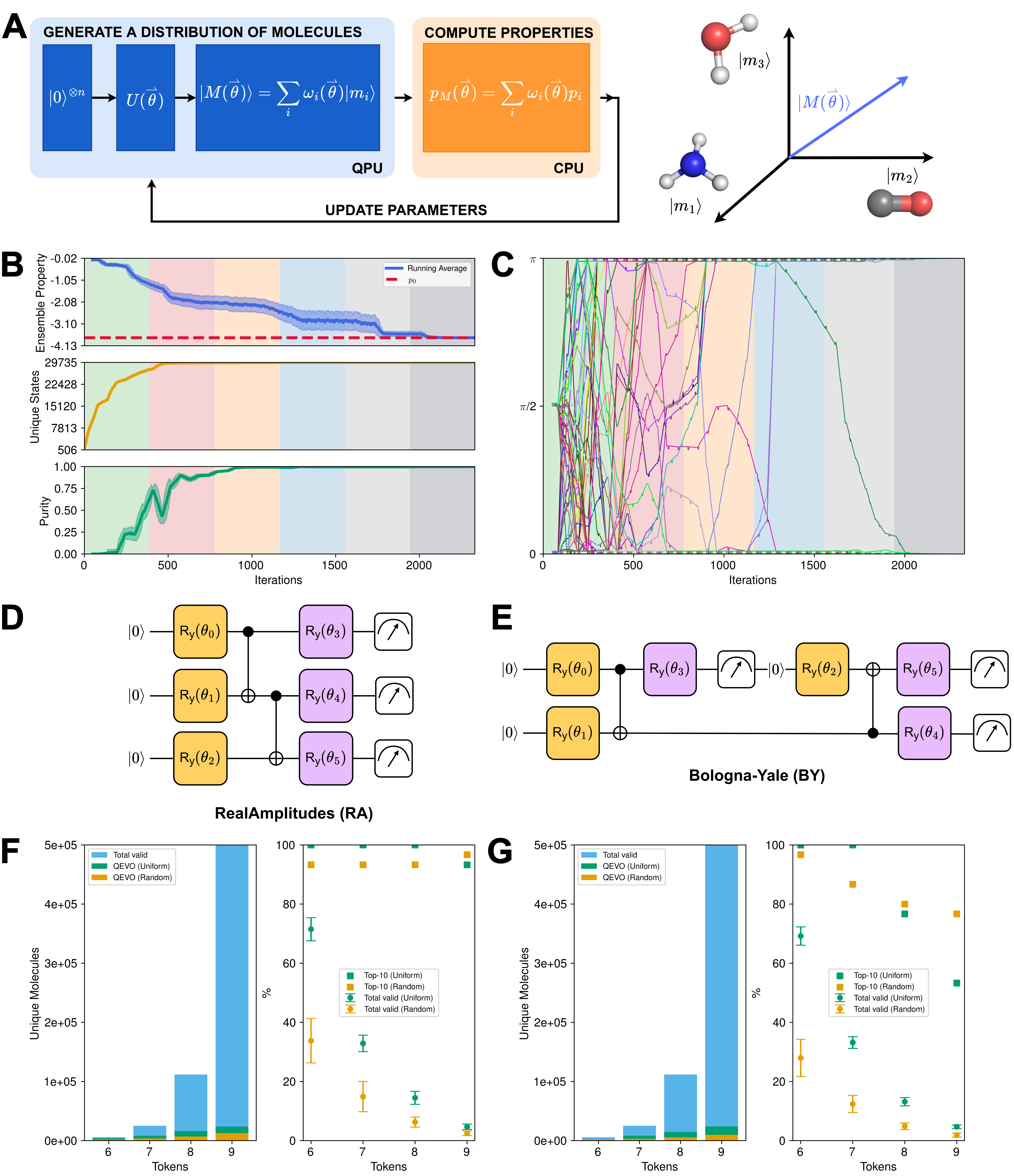}
    \caption{\textbf{The Quantum Ensemble Variational Optimization (QEVO) Algorithm.} 
\textbf{A.} Schematic of the QEVO workflow. Variational parameters of a unitary transformation are optimized to generate quantum states for sampling candidate molecules. A classical computer evaluates the ensemble-averaged property value, \( p_M \), and guides parameter updates. 
\textbf{B.} Convergence behavior of QEVO during inverse design of molecules with up to 9 ASCII tokens targeting minimization of the plogP value. Shown are the ensemble-averaged property value (blue solid line), the best property value in the full reference space, \( p_0 \) (red dashed line), the cumulative number of unique molecules sampled (orange solid line), and the purity of the quantum ensemble (green solid line). Shaded areas denote running standard deviations over 50 independent simulations. 
\textbf{C.} Colored solid lines show the time evolution of ansatz parameters averaged across simulations. Panels \textbf{B} and \textbf{C} are segmented into six optimization windows, indicated by different background colors, to illustrate distinct phases during training.
\textbf{D–E.} Structure of the RealAmplitudes (RA, \textbf{D}) and Bologna-Yale (BY, \textbf{E}) ansätze. The BY ansatz is a holographic, qubit-efficient variant that uses iterative measurement and initialization of two qubits. 
\textbf{F–G.} Comparison of QEVO’s sampling behavior across different ansätze and initialization schemes. The total number of valid molecules in each chemical space is shown (blue histograms), alongside the number of unique molecules sampled using the RA (\textbf{F}) or BY (\textbf{G}) ansatz with uniform (orange) or random (green) initialization. Circular markers denote the average fraction of unique molecules explored under each scheme, and square markers indicate the success rate of generating at least one top-10-ranked solution. Results are averaged over 30 independent runs for each of the 6-, 7-, 8-, and 9-token combinatorial spaces, with 1,024 binary strings sampled per optimization step.
}
    \label{panel1}
\end{figure}

As a benchmark study, we first considered a tractable example where the full set of possible molecules could be explicitly characterized. Specifically, we generated all combinations of 6, 7, 8, and 9 ASCII tokens defining molecules drawn from a dictionary of \(2^3\) elements (see Sec.~\ref{methods} for details on the vocabulary), corresponding to combinatorial spaces of dimensionality \(2^{18}\), \(2^{21}\), \(2^{24}\), and \(2^{27}\), respectively. Due to degeneracies in the mapping between token combinations and valid molecular representations, the number of unique, meaningful combinations was reduced to 5,790; 25,218; 111,711; and 504,183, respectively (blue bars in Fig.~\ref{panel1}, F and G). However, it is important to highlight that this reduction does not affect the combinatorial complexity of the problems, which is still equal to the number of total combinations (hereafter called {\em full reference chemical space}).

Figure~\ref{panel1}B illustrates the performance of QEVO in a representative simulation using the RealAmplitudes (RA) ansatz. QEVO successfully identifies the solution with minimum plogP after 2,332 optimization steps, with 1,024 binary strings sampled per step. Remarkably, convergence is achieved after sampling only 29,735 unique combinations—corresponding to just 6\% of all unique molecules in the full chemical space.

Upon initialization, the ansatz produces a uniform superposition, leading to uniform sampling of binary strings. As expected, the cumulative number of unique sampled molecules grows rapidly during the early stages of optimization (Fig.~\ref{panel1}B). Over time, as the optimization progresses, the purity of the ensemble increases and the quantum state \( |M(\vec{\theta})\rangle \) gradually aligns with the binary string that minimizes the loss function. In this case, the final solution corresponds to \textit{n}-nonane, which exhibits the lowest loss with a predicted plogP of $-3.7569$.

QEVO's efficiency arises from the rapid convergence of the ansatz parameters (Fig.~\ref{panel1}C), which evolve quickly to reduce the loss, despite initially generating poorly scoring candidates. The increasing purity of the quantum state is reflected in the convergence of the ansatz rotation angles toward either \(0\) or \(\pi\), corresponding to pure basis states \( |0\rangle \) and \( |1\rangle \) for each qubit (Fig.~\ref{panel1}C).

We compared the performance of QEVO using both the RealAmplitudes (RA) ansatz (Fig.~\ref{panel1}D) and its holographic implementation~\cite{lyu2024holographic}—hereafter referred to as the Bologna-Yale (BY) ansatz—which operates by iteratively measuring and resetting two qubits (Fig.~\ref{panel1}E). Both ansätze were tested under two initialization schemes: uniform superpositions and randomly sampled parameters. 

Across all configurations, QEVO demonstrated strong performance, consistently identifying high-quality molecular candidates while sampling only a small fraction of the full solution space (see Sec.~\ref{sec:simulation_details}). As anticipated, the RA ansatz outperforms the BY variant, primarily due to the trade-off introduced by the latter's qubit-efficient design.

In simulations over the \(2^{27}\)-dimensional chemical space, the RA and BY ansätze, under uniform initialization, yield solutions ranked in the top 10 of the reference set in 93\% and 53\% of the runs, respectively (see Sec.~\ref{sec:simulation_details}). When initialized randomly, performance improves: RA achieves a 97\% success rate, while BY reaches 77\%. Given that the BY ansatz represents a 27-qubit problem with only 2 qubits, these results are particularly noteworthy. They highlight QEVO's resource-adaptive nature and its capacity to operate effectively within the physical and coherence-time constraints of near-term quantum hardware.

On average, uniform initialization yields better results than random initialization in smaller combinatorial spaces, though it requires more extensive exploration (Fig.~\ref{panel1}F,G). Interestingly, QEVO's performance improves as the size of the reference space increases, and the impact of the initialization scheme diminishes. This trend stems from the fixed number of binary strings sampled per iteration: as the size of the solution space grows, the relative proportion of unique states explored naturally decreases. For example, using the RA ansatz with uniform initialization, QEVO explores on average \(72 \pm 4\%\) of unique valid states in the \(2^{18}\) space, compared to only \(5 \pm 1\%\) in the \(2^{27}\) space (Fig.~\ref{panel1}F). With random initialization, these exploration rates drop to \(34 \pm 8\%\) and \(3 \pm 1\%\), respectively. These trends illustrate QEVO's ability to maintain strong performance while requiring increasingly limited exploration in larger search spaces.

\FloatBarrier
\subsection{Stairways to New Drugs}

We further evaluated QEVO in the context of drug-like molecule design (see Sec. \ref{methods} for details on the loss function). When applied to the \(2^{27}\)-dimensional combinatorial space using the RA ansatz with uniform initialization, QEVO explored on average \(7 \pm 1\%\) of the unique valid chemical space and successfully identified candidates ranked in the top 10 of the reference set in 90\% of simulations (Fig.~\ref{trend_drug_design}). These results are remarkable given the complexity of the underlying property landscape. 

\begin{figure}[!ht]
    \centering
    \includegraphics[width=1.\linewidth]{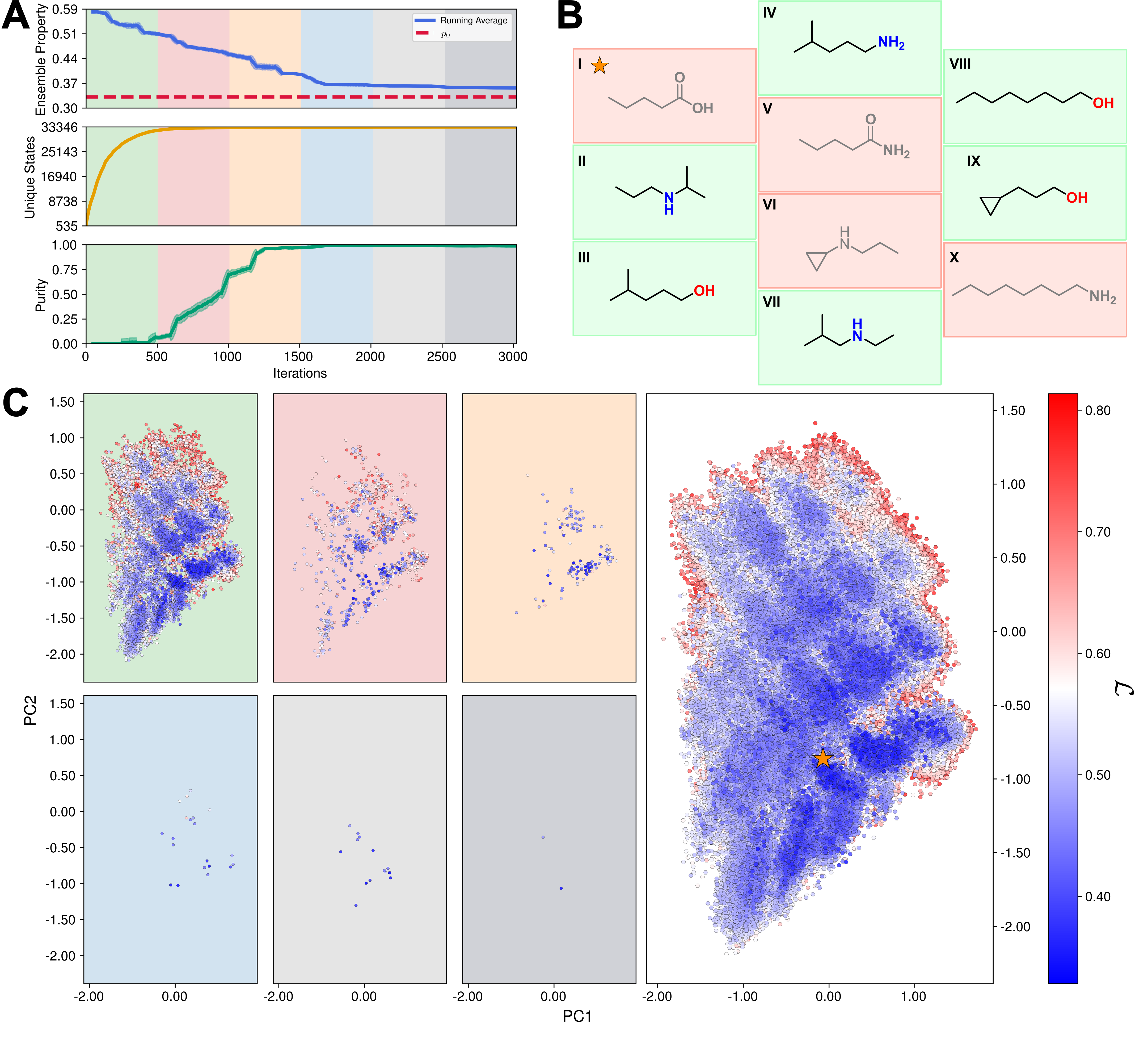}
    \caption{\textbf{Drug Design with QEVO.} Results from a representative QEVO simulation (out of 30 independent runs) for the inverse design of drug-like molecules composed of up to 9 ASCII tokens, using the RealAmplitudes ansatz with uniform initialization.
\textbf{A.} Convergence of the running average (over 50 samples) of the ensemble property value (blue solid line) toward the best value \( p_0 \) (red dashed line) within the full reference space. Also shown are the cumulative number of unique molecules sampled (orange solid line) and the evolution of the quantum ensemble's purity (green solid line). The optimization process is segmented into six stages, indicated by different background colors.
\textbf{B.} Top-10 molecules from the reference space are listed, and those sampled during the QEVO run reported in panel \textbf{A} are highlighted in green.
\textbf{C.} Projection of newly generated unique molecules onto the principal component space computed from the full reference set (right subpanel). Each point corresponds to a molecule sampled during a given optimization window (as defined in panel \textbf{A}) and is colored according to its associated loss function value. The global optimum, identified in the reference space, is marked with an orange star.
}
    \label{panel2}
\end{figure}

This difficulty is reflected in the nature of the final quantum state: unlike in the plogP optimization tasks, the final state here is never pure and does not collapse to pentanoic acid—the global optimum in this design problem (Fig.~\ref{panel2}B). In contrast, for plogP minimization the final state converges to a pure solution in 30\% of runs when using the RA ansatz with uniform initialization, indicating a less rugged optimization surface in that case.

As shown in Fig.~\ref{panel2} for a representative simulation, even when QEVO does not converge to the global optimum, it effectively optimizes the quantum state to generate high-quality suboptimal ensembles. In this simulation, the final ensemble reaches a purity of 99\% and explores 7\% of all valid solutions (Fig.~\ref{panel2}A). Despite not collapsing to the optimal state, QEVO samples 60\% of the top-ranked solutions throughout the course of optimization (Fig.~\ref{panel2}B). This behavior is further illustrated by a time-resolved principal component analysis (PCA) performed over the sampled molecules during the simulation (see Sec.~\ref{methods} for computational details). The PCA projection was computed using principal components derived from the full reference space (Fig.~\ref{panel2}C, right panel). The new molecules generated by QEVO were then projected onto this space to visualize the trajectory of the optimization process. 

Consistent with uniform initialization, the algorithm initially samples a wide distribution of molecular states during the first sixth of the optimization (green subpanel, Fig.~\ref{panel3}A,B). As optimization proceeds, the distribution narrows and concentrates around high-reward regions, ultimately approaching the vicinity of the optimal state (orange star, Fig.~\ref{panel2}C).

\subsection{JAK2 Inhibition}

While exploring a combinatorial space of \(2^{27}\) molecular candidates is already computationally demanding, it still falls short of the scale required for many real-world drug discovery applications. Approved pharmaceuticals often consist of complex structures containing tens of atoms, which correspond to much longer ASCII strings than the 9-token examples considered thus far. A case in point is ruxolitinib~\cite{ruxolitinib}, a clinically approved Janus kinase 2 (JAK2) inhibitor~\cite{ruxolitinib_crystal}.

JAK2 mutations are a key driver of myelofibrosis, a rare hematological malignancy marked by the abnormal and excessive proliferation of blood cells~\cite{jak2_review}. Developing selective JAK2 inhibitors remains particularly challenging due to the need to distinguish JAK2 from structurally similar off-target kinases—most notably, lymphocyte-specific protein tyrosine kinase (LCK), which is not implicated in myelofibrosis~\cite{LCKvsJAK2}. The structural similarity between these kinases imposes a high selectivity barrier, making JAK2 inhibition an ideal benchmark for evaluating molecular design algorithms. In this context, QEVO offers a compelling testbed, as it must optimize candidate molecules not only for potency against JAK2 but also for selectivity over competing targets within a large and complex chemical space.

Prompted by these findings, we extended QEVO to the inverse design of novel JAK2 inhibitors using 40-token-long ASCII strings and a vocabulary of \(2^4\) tokens (see Sec.~\ref{methods} for details on the vocabulary), yielding a combinatorial design space of \(2^{160}\) possible solutions, for which verifying the density of valid states is already computationally unfeasible. The choice of 40 tokens reflects the encoded size of ruxolitinib, while allowing additional flexibility to explore chemically diverse candidates.

\begin{figure}[h]
    \centering
    \includegraphics[width=1.0\linewidth]{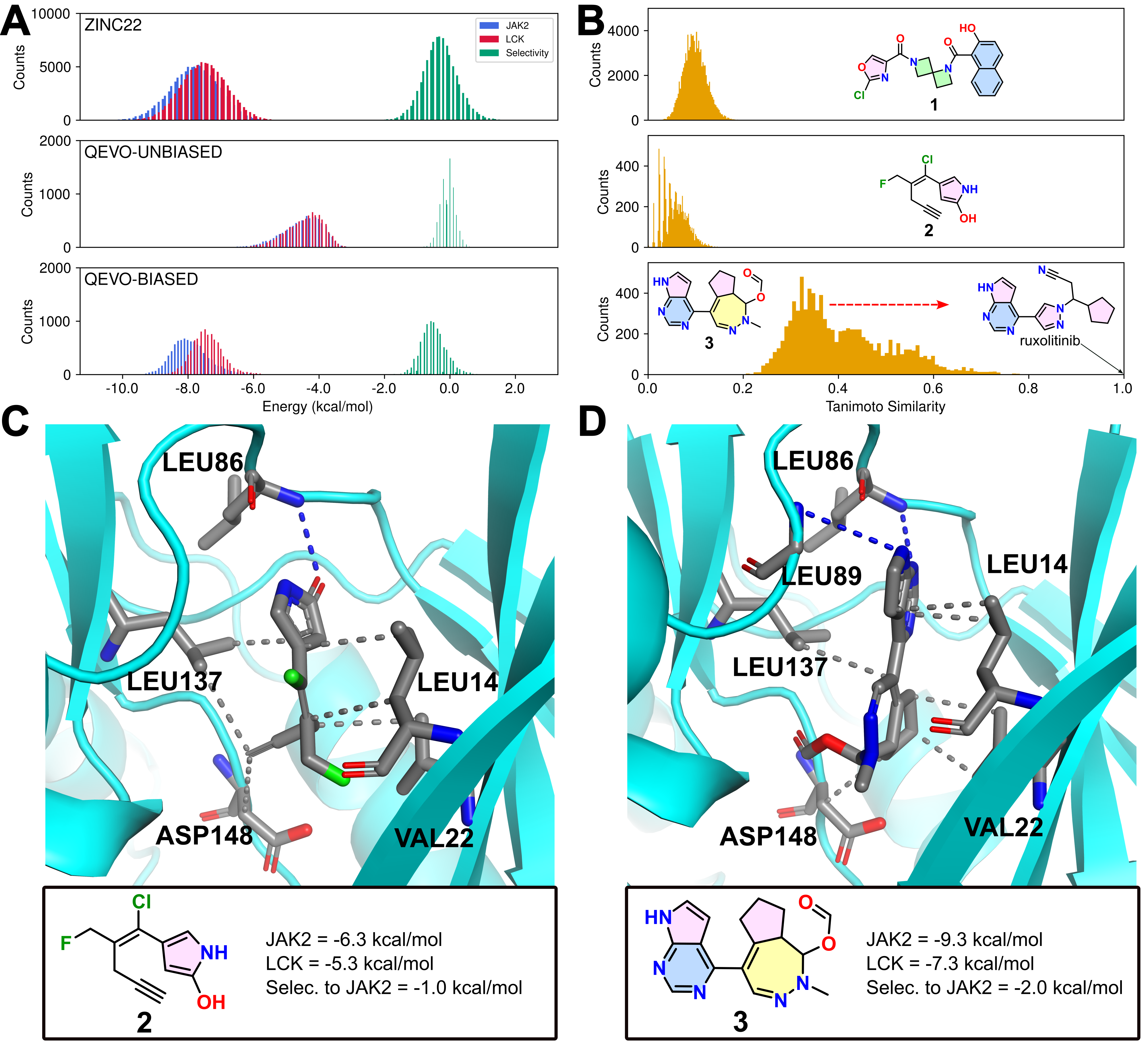}
    \caption{\textbf{Design of a Selective JAK2 Inhibitor.} 
\textbf{A.} Distributions of docking energies for candidate molecules against JAK2 (blue), LCK (crimson), and the binding energy difference favoring JAK2 over LCK (green). Results are shown for 100,000 randomly sampled compounds from the ZINC22 database, and the top 10,000 candidates generated by QEVO under unbiased and similarity-biased optimization schemes.
\textbf{B.} Distributions of chemical similarity to the reference inhibitor (ruxolitinib) across the three candidate sets in panel \textbf{A} are reported. In the insets, the fittest candidates are reported for ZINC and the two QEVO simulations. 
\textbf{C–D.} Docking poses of the top-scoring QEVO-generated molecules from the unbiased (\textbf{C}) and similarity-biased (\textbf{D}) design runs. Hydrogen bonds and hydrophobic interactions are indicated by blue and gray dashed lines, respectively.
}
\label{panel3}
\end{figure}

We first targeted the generation of drug-like molecules, as described earlier, and selected the top 10,000 candidates generated by QEVO for molecular docking. While QEVO successfully generated molecules with strong drug-like properties (QEVO-UNBIASED generation, Fig.~\ref{mols_PANEL_unbiased}), their selectivity for JAK2 over LCK remained limited.
To assess this, we compared the QEVO-generated molecules to a baseline of 100,000 randomly sampled compounds from the ZINC22 database~\cite{zinc22}. QEVO’s candidates scored worse in terms of binding selectivity: the best-performing molecule from ZINC22 exhibited a binding energy difference between JAK2 and LCK of \(-2.8\) kcal/mol, while the top QEVO-derived compound showed only \(-1.0\) kcal/mol (species \textbf{1} and \textbf{2}, respectively, Fig.~\ref{panel3}B).

Interestingly, species \textbf{2}, designed by QEVO, acts as a type I kinase inhibitor, similar to ruxolitinib~\cite{ruxolitinib_crystal}, forming hydrogen bonds with the hinge region via residue LEU86 (Fig.~\ref{panel3}C)~\cite{inhibitors_types}. In contrast, species \textbf{1} from ZINC22 is a type V inhibitor, interacting with multiple binding sites on JAK2 (Fig.~\ref{docking_ZINC22}).

Relying solely on docking scores may not be sufficient for optimal drug design. Although ruxolitinib exhibits modest selectivity for JAK2—with a differential binding energy of only 0.6 kcal/mol as determined by our molecular docking studies—its FDA approval underscores its clinical relevance and provides valuable structural insights warranting further exploration. To this end, we evaluated the chemical similarity of each candidate molecule to ruxolitinib (see Sec.~\ref{methods} for computational details). Neither molecules from the ZINC22 database nor those generated by QEVO approached close similarity to the reference compound (Fig.~\ref{panel3}B).  Consequently, we performed an additional QEVO experiment in which molecules were positively rewarded based on their similarity to ruxolitinib. To enhance convergence, the ansatz parameters were initialized so that the quantum state \(|M(\vec{\theta})\rangle\) aligned with the encoding of ruxolitinib—effectively starting the optimization within a known, suboptimal subspace.

Docking of the top 10,000 QEVO-generated candidates from this biased run confirmed improved selectivity: their binding affinities to JAK2 and LCK more closely resembled those of molecules in ZINC22, and their average similarity to ruxolitinib increased (QEVO-BIASED generation, Fig.~\ref{panel3}A,B). Specifically, the separation between the median affinities for JAK2 and LCK was 0.3 kcal/mol for ZINC22, 0.1 kcal/mol for unbiased QEVO, and increased to 0.5 kcal/mol in the biased QEVO simulation (Fig.~\ref{panel3}A).
The top candidate from the biased QEVO run, species \textbf{3} (Figs.~\ref{panel3}B and \ref{mols_PANEL_biased}), exhibits a 2.0 kcal/mol stronger binding to JAK2 over LCK and shares 31\% similarity with ruxolitinib. Like species \textbf{2}, species \textbf{3} forms hydrogen bonds with the hinge region via LEU86 and LEU89 residues (Fig.~\ref{panel3}D), alongside hydrophobic interactions at multiple protein sites, indicative of a Type I kinase inhibitor.

These findings highlight the capabilities of QEVO for molecular design, enabling rapid identification of promising anticancer candidates without reliance on training of neural networks. Furthermore, by adjusting ansatz initialization strategies, QEVO offers flexible control over chemical space exploration, making it a versatile tool for diverse inverse design challenges.

\FloatBarrier
\section{Conclusions}\label{discussion}

Inverse design in chemistry—and combinatorial optimization more broadly—remains a significant challenge due to the exponential size of the solution spaces, which precludes exhaustive enumeration and characterization. 
In this work, we introduced QEVO, a variational quantum algorithm that optimizes ensemble averages computed from molecules sampled from a distribution function defined by measurements of a quantum ansatz in the computational basis. We demonstrated QEVO as applied to molecular design. Our simulations highlight its effectiveness to generate high-quality molecular candidates.

We find that QEVO operates efficiently using either constant-depth or constant-qubit RealAmplitudes ansätze, requiring modest quantum resources. This makes it well-suited for timely, resource-constrained applications as illustrated for the design of novel anticancer drugs.

Overall, QEVO represents a promising algorithm that leverages quantum superposition directly, circumventing the need for explicit Hamiltonian embedding. This approach opens a novel computational paradigm for tackling large-scale combinatorial problems.

\FloatBarrier
\section{Methods}\label{methods}

\subsection{The Mapping Scheme}

Molecules are encoded using the SELFIES representation~\cite{SELFIES}, which expresses each molecule as a sequence of discrete tokens \( t_i \). To interface this representation with a quantum system, SELFIES strings are bijectively mapped to quantum states. Each token \( t_i \) is assigned a binary string \( |q_i\rangle \), represented in the computational basis \( \mathcal{C} = \{ |0\rangle, |1\rangle \} \). To encode a vocabulary of \( 2^n \) unique SELFIES tokens, \( n \) qubits are required:

\begin{equation}
    |q_i\rangle = \bigotimes_{j=1}^n |c_j\rangle, \quad \text{where} \quad |c_j\rangle \in \mathcal{C}.
    \label{eq:token_mapping}
\end{equation}

Given the binary-encoded SELFIES tokens \( |q_i\rangle \in \mathcal{Q} \), where \( \mathcal{Q} \) is the token Hilbert space, a molecule \( m_i \) composed of up to \( k \) tokens is mapped to a unique quantum state \( |m_i\rangle \) by concatenating the binary encodings of its constituent tokens:

\begin{equation}
    |m_i\rangle = \bigotimes_{j=1}^k |q_j\rangle, \quad \text{where} \quad |q_j\rangle \in \mathcal{Q}.
    \label{eq:molecular_mapping}
\end{equation}

Since both \( \mathcal{C} \) and \( \mathcal{Q} \) are orthonormal bases, the resulting molecular space \( \mathcal{M} \), composed of all \( |m_i\rangle \), also forms an orthonormal basis. When constructed using all possible token combinations of length up to \( k \) from a fixed vocabulary, \( \mathcal{M} \) spans the entire relevant chemical design space.

\subsection{Purity of the Ensemble}

The purity \( P(\vec{\theta}) \) of the ensemble \( |M(\vec{\theta})\rangle \), generated by measurements of the quantum state parameterized by \( \vec{\theta} \), quantifies the diversity of molecular candidates sampled from the chemical space. It is defined as:

\begin{equation}
    P(\vec{\theta}) = 1 - \frac{|\mathrm{set}(|M\rangle)|}{x_e},
    \label{eq:purity_metric}
\end{equation}
where \( x_e \) is the total number of samples (i.e., ensemble size), and \( |\mathrm{set}(|M\rangle)| \) denotes the number of unique quantum states (Pauli words) observed in the ensemble.

A purity value close to 1 indicates that the ensemble is dominated by a small number of distinct molecules—suggesting convergence toward a specific molecular candidate—while values near 0 indicate a broad, highly diverse ensemble of molecules.

\subsection{Ansatz Implementation}

We implemented two variational ansätze: the RealAmplitudes (RA) ansatz and the Bologna-Yale (BY) ansatz. 

The RA ansatz applies a linearly entangled unitary transformation \( U(\vec{\theta}) \) composed of real amplitude rotations to the vacuum state. The circuit features alternating layers of single-qubit parameterized rotations and linear entangling gates. This structure ensures constant circuit depth and a number of trainable parameters \( \vec{\theta} \) that scales linearly with the number of qubits.

The BY ansatz is a holographic implementation~\cite{lyu2024holographic} of the RA circuit. Although multiple holographic implementations are possible, in the present work we chose to use only two physical qubits, which are repurposed by iterative measurement and reinitialization.  As a result, the BY ansatz maintains a constant number of qubits and circuit depth, while the number of variational parameters \( \vec{\theta} \) scales linearly with the size of the target mapped state.
This trade-off allows the BY ansatz to compress high-dimensional quantum states into hardware-constrained systems, enabling efficient optimization even on near-term quantum processors.

\subsection{Initialization Mode}

To ensure a representative initial sampling of the solution space, the variational quantum circuit \( U(\vec{\theta}) \) is initialized in a uniform superposition over all possible basis states. Applying a Hadamard gate to each qubit,
\begin{equation}
    H|0\rangle = \frac{1}{\sqrt{2}}(|0\rangle + |1\rangle),
    \label{eq:superpos}
\end{equation}
we obtain the state of the \( n \)-qubit register:
\begin{equation}
    H^{\otimes n}|0\rangle^{\otimes n} = \frac{1}{\sqrt{2^n}} \sum_{x \in \{0,1\}^n} |x\rangle,
\end{equation}
as the hyperdiagonal of the molecular Hilbert space \( \mathcal{M} \). This uniform initialization ensures that the quantum state \( |M(\vec{\theta})\rangle \) initially contains equal probability amplitude over all possible molecular configurations encoded in the computational basis.

The RA and BY ansätze consist of two layers of parameterized \( R_y(\theta) \) rotations separated by one (RA) or two (BY) entangling gates. Since a Hadamard gate is equivalent to a \( R_y(\pi/2) \) rotation up to a phase in the \( XZ \)-plane, we initialize the ansatz to approximate this behavior.
Specifically, the angles of the \( R_y \) gates preceding the entangling operations are set to 0:
\begin{equation}
    R_y(0)|0\rangle = |0\rangle,
\end{equation}
ensuring that all control qubits in subsequent entangling gates (e.g., CNOT) remain in the \( |0\rangle \) state
\begin{equation}
    \text{CNOT}_{i,j}|00\rangle = |00\rangle.
\end{equation}
Following entanglement, the second set of \( R_y \) gates is initialized to \( \frac{\pi}{2} \), yielding
\begin{equation}
    R_y\left(\frac{\pi}{2}\right)|0\rangle = \frac{1}{\sqrt{2}}(|0\rangle + |1\rangle),
\end{equation}
which completes the transformation of the vacuum state into a uniform superposition. Thus, \( U(\vec{\theta})|0\rangle^{\otimes n} = H^{\otimes n}|0\rangle^{\otimes n} \), ensures that the initial variational state spans the entire solution space for unbiased initial exploration of the chemical space.

\subsection{Loss Functions}

Here, we define the loss functions used for quantum optimization. These functions are evaluated for each sampled molecule \( m_i \), according to chemical properties, or structural similarity to a reference molecular structure.

\subsubsection*{plogP Loss}  
The first objective focuses on the optimization of the negative logarithm of the octanol-water partition coefficient (plogP)~\cite{logP}, which is widely used to assess molecular solubility and permeability. The loss function is defined as:
\begin{equation}
\mathcal{L}(m_i) = 
\begin{cases}
  \text{plogP}(m_i) & \text{if $m_i$ is a valid molecule}, \\
  1.0 & \text{otherwise},
\end{cases}
\end{equation}
where a penalty of 1.0 is applied to invalid molecular encodings.

\subsubsection*{Drug Design Loss}  
For the design of drug-like molecules, we define a composite loss function that integrates three terms:
\begin{itemize}
  \item \( a(m_i) \): A modified QED-based term penalizing low drug-likeness and large ring sizes~\cite{qed},
  \item \( b(m_i) \): A normalized synthetic accessibility score (SAS)~\cite{sas},
  \item \( c(m_i) \): A dissimilarity score based on 1 minus the Tanimoto similarity to a reference molecule \( m_{\text{ref}} \)~\cite{tanimoto}.
\end{itemize}

These components are defined as follows:
\begin{align}
    a(m_i) &= 
    \begin{cases}
        1.0 - \text{QED}(m_i) & \text{if rings with more than 7 atoms in $m_i$}, \\
        1.2 - \text{QED}(m_i) & \text{otherwise},
    \end{cases} \quad a \in [0.0, 1.2], \\
    b(m_i) &= \frac{\text{SAS}(m_i)}{10.0}, \quad b \in [0.1, 1.0], \\
    c(m_i) &= 1.0 - \text{S}(m_i, m_{\text{ref}}), \quad c \in [0.0, 1.0].
\end{align}

The final weighted loss is given by:
\begin{equation}
\mathcal{L}(m_i) = 
\begin{cases}
  \dfrac{\alpha \cdot a(m_i) + \beta \cdot b(m_i) + \gamma \cdot c(m_i)}{\alpha + \beta + \gamma} & \text{if $m_i$ is a valid molecule}, \\
  1.0 & \text{otherwise},
\end{cases}
\end{equation}
where \( \alpha, \beta, \gamma \) are user-defined weights that control the relative importance of each objective.

All molecular properties (QED, SAS, plogP) and similarity scores are computed using the \texttt{RDKit} Python package~\cite{rdkit}, after converting SELFIES representations into SMILES strings~\cite{smiles1,smiles2,smiles3}.

\subsection{Postprocessing}

Principal Component Analysis (PCA) was performed using the \texttt{Scikit-Learn} Python package~\cite{scikit-learn}, with molecules encoded as 1,024-bit \texttt{MorganFingerprints} (radius = 3), as implemented in \texttt{RDKit}~\cite{rdkit}.

All molecular docking calculations were carried out using the \texttt{Dockstring} Python package~\cite{dockstring}. Docking poses were analyzed with the Protein–Ligand Interaction Profiler (PLIP) web server~\cite{plip}, and visualizations were rendered using \texttt{PyMOL}~\cite{pymol}.

\subsection{Simulation Details}

Quantum simulations were conducted in the \texttt{Qiskit} framework (v1.3.1) using the \texttt{AerSimulator} backend (qiskit-aer v0.15.1) with the \texttt{matrix\_product\_state} simulation method~\cite{qiskit2024}.

\subsubsection*{Token-Based Simulations (6–9 tokens)}  
For simulations involving molecular encodings of 6 to 9 SELFIES tokens, we used the token-to-qubit mapping defined in Table~\ref{mapping1}. Each optimization iteration sampled 1,024 Pauli strings. The variational parameters were optimized using the Implicit Filtering (ImFil) algorithm from \texttt{QiskitAlgorithms} (v0.3.1). Drug-like molecule generation was guided by the weighted loss function described in Sec.~\ref{methods}, with coefficients set to \( \alpha = 2 \), \( \beta = 1 \), and \( \gamma = 0 \).

\subsubsection*{Large-Scale Simulations (40 tokens)}  
For higher-dimensional simulations (e.g., involving 40-token encodings), the mapping scheme in Table~\ref{mapping2} was employed. Each iteration involved sampling 10,240 Pauli strings. Optimization was performed using the Simultaneous Perturbation Stochastic Approximation (SPSA) algorithm with 50 resampling evaluations per step, also from \texttt{QiskitAlgorithms}. In these simulations, the loss function included a similarity constraint to ruxolitinib, using coefficients \( \alpha = 2 \), \( \beta = 1 \), and \( \gamma = 2 \), where \( \gamma \) weights the Tanimoto similarity to the reference molecule.

\begin{table}[!htb]
\centering
\caption{$2^3$ SELFIES-Pauli string mapping.} \label{mapping1}
\begin{tabular}{@{}ll@{}}
\toprule
SELFIES token & Pauli string\\
\midrule
\texttt{[C]} & $|000\rangle$\\
\texttt{[O]} & $|001\rangle$\\
\texttt{[N]} & $|010\rangle$\\
\texttt{[F]} & $|011\rangle$\\
\texttt{[=C]} & $|100\rangle$\\
\texttt{[\#N]} & $|101\rangle$\\
\texttt{[Ring1]} & $|110\rangle$\\
\texttt{[Branch1]} & $|111\rangle$\\
\bottomrule
\end{tabular}
\end{table}

\begin{table}[ht]
\centering
\caption{$2^4$ SELFIES-Pauli string mapping.} \label{mapping2}%
\begin{tabular}{@{}ll@{}} 
\toprule
SELFIES token & Pauli string\\
\midrule
\texttt{[C]} & $|0000\rangle$ \\
\texttt{[=C]} & $|1000\rangle$ \\
\texttt{[\#C]} & $|0100\rangle$ \\
\texttt{[O]} & $|0010\rangle$ \\
\texttt{[=O]} & $|0001\rangle$ \\
\texttt{[N]} & $|1100\rangle$ \\
\texttt{[=N]} & $|0011\rangle$ \\
\texttt{[\#N]} & $|0110\rangle$ \\
\texttt{[F]} & $|1001\rangle$ \\
\texttt{[Cl]} & $|1010\rangle$ \\
\texttt{[Ring1]} & $|0101\rangle$ \\
\texttt{[Ring2]} & $|1110\rangle$ \\
\texttt{[Branch1]} & $|0111\rangle$ \\
\texttt{[=Branch1]} & $|1101\rangle$ \\
\texttt{[Branch2]} & $|1011\rangle$ \\
\texttt{[=Branch2]} & $|1111\rangle$ \\
\bottomrule
\end{tabular}
\end{table}

\FloatBarrier
\section{Funding}
F.C. acknowledges the ``Ing. Luciano Toso Montanari" Foundation for financially supporting his secondment at Yale University (U.S.A.) in the group of Professor Victor S. Batista, and partial support from the National Science Foundation Engines Development Award: Advancing Quantum Technologies (CT) under Award Number 2302908.
V.S.B. acknowledges support from the NSF Center for Quantum Dynamics on Modular Quantum Devices (CQD-MQD) under grant number 2124511.

\section{Conflict of interest}
F.C., D.G.A.C., I.R., and V.S.B. are listed as inventors of a patent application on the QEVO algorithm.

\section{Author contributions}
F.C. conceived the idea of QEVO, wrote the Python code, performed experiments, and drafted the manuscript. D.G.A.C. validated the results and contributed to software development. All authors discussed the results and revised the manuscript. I.R. and V.S.B. supervised the work and provided financial support.

\newpage
\begin{appendices}
\renewcommand{\thefigure}{A.\arabic{figure}}
\setcounter{figure}{0}
\renewcommand{\thetable}{A.\arabic{table}}
\setcounter{table}{0}
\renewcommand{\thealgorithm}{A.\arabic{algorithm}}
\setcounter{algorithm}{0}
\FloatBarrier
\section{Supporting Data}\label{appendix}

\subsection{Pseudocode of the Quantum Ensemble Variational Optimizer}\label{sec7}

\begin{algorithm}[ht]
\caption{Pseudo-code for the QEVO algorithm}\label{alg:QEVO}
\begin{algorithmic}
\Require $U(\vec{\theta})$, $t_i \mapsto |q_i\rangle$, $\mathcal{L}$, $\lambda$.
\Ensure $\Delta\mathcal{L} \leq \varepsilon$
\State Initialize $|M(\vec{\theta})\rangle = U(\vec{\theta}) |0\rangle^{\otimes n}$ with $\theta \in [0, \pi]$
\While{not all converged $\vec{\theta}$}
    \State Measure $|M(\vec{\theta})\rangle$
    \State Compute the property value $p_M = \sum_i \omega_i p_i$ 
    \State Compute the gradient of $\vec{\theta}$ with respect to $p_M$
    \State Update parameters $\vec{\theta}$
\EndWhile
\State Return the final optimized parameters $\vec{\theta}$.
\end{algorithmic}
\end{algorithm}

\clearpage
\pagebreak

\subsection{Simulations data} \label{sec:simulation_details}

\begin{table}[ht!]\centering
\caption{Details on the QEVO simulations using 6 SELFIES tokens, uniform initialization and the Bologna-Yale ansatz. The number of the repetition, the SMILES string of the best state generated, its loss value, $\mathcal{L}$, if QEVO generates at least one state belonging to the top-10 loss value positions as characterized in the reference solution space, and the number of unique states generated during the current simulation are reported.}%
\begin{tabular}{@{}lllll@{}}
\toprule
Repetition & Best state           & $\mathcal{L}^1$  & In the top-10? & Unique states sampled \\
\midrule
1  & \texttt{CCCCCC}  & -2.5866 & True & 4,577 \\
2  & \texttt{CCCCCC}  & -2.5866 & True & 3,882 \\
3  & \texttt{CCCCCC}  & -2.5866 & True & 4,196 \\
4  & \texttt{CCCCCC}  & -2.5866 & True & 4,245 \\
5  & \texttt{CCCCCC}  & -2.5866 & True & 4,094 \\
6  & \texttt{CCCCCC}  & -2.5866 & True & 3,805 \\
7  & \texttt{CCCCCC}  & -2.5866 & True & 4,071 \\
8  & \texttt{CCCCCC}  & -2.5866 & True & 3,817 \\
9  & \texttt{CCCCCC}  & -2.5866 & True & 3,911 \\
10 & \texttt{CCC=CCC} & -2.3626 & True & 4,016 \\
11 & \texttt{CCCCCC}  & -2.5866 & True & 4,211 \\
12 & \texttt{CCCCC}   & -2.1965 & True & 3,932 \\
13 & \texttt{CCCCCC}  & -2.5866 & True & 3,737 \\
14 & \texttt{CCCCCC}  & -2.5866 & True & 3,992 \\
15 & \texttt{CCCCCC}  & -2.5866 & True & 4,021 \\
16 & \texttt{CCCCCC}  & -2.5866 & True & 3,858 \\
17 & \texttt{CCCCCC}  & -2.5866 & True & 3,972 \\
18 & \texttt{CCCCCC}  & -2.5866 & True & 3,859 \\
19 & \texttt{CCCCCC}  & -2.5866 & True & 3,933 \\
20 & \texttt{CCCCCC}  & -2.5866 & True & 3,926 \\
21 & \texttt{CCCCCC}  & -2.5866 & True & 4,081 \\
22 & \texttt{CCCCCC}  & -2.5866 & True & 3,797 \\
23 & \texttt{CCCCCC}  & -2.5866 & True & 3,850 \\
24 & \texttt{CCCCCC}  & -2.5866 & True & 3,931 \\
25 & \texttt{CCCCCC}  & -2.5866 & True & 3,944 \\
26 & \texttt{CCCCCC}  & -2.5866 & True & 4,130 \\
27 & \texttt{CCCCCC}  & -2.5866 & True & 3,899 \\
28 & \texttt{CCCCCC}  & -2.5866 & True & 4,301 \\
29 & \texttt{CCCCCC}  & -2.5866 & True & 4,082 \\
30 & \texttt{CCCCCC}  & -2.5866 & True & 4,141 \\
\bottomrule
\multicolumn{5}{l}{$^1$plogP.} \\
\end{tabular}
\end{table}

\begin{table}[ht!]\centering
\caption{Details on the QEVO simulations using 6 SELFIES tokens, uniform initialization and the RealAmplitudes ansatz. The number of the repetition, the SMILES string of the best state generated, its loss value, $\mathcal{L}$, if QEVO generates at least one state belonging to the top-10 loss value positions as characterized in the reference solution space, and the number of unique states generated during the current simulation are reported.}%
\begin{tabular}{@{}lllll@{}}
\toprule
Repetition & Best state           & $\mathcal{L}^1$  & In the top-10? & Unique states sampled \\
\midrule
1  & \texttt{C=CCCCC}  & -2.3626 & True & 4,039 \\
2  & \texttt{CCCCCC}   & -2.5866 & True & 4,087 \\
3  & \texttt{CCCCCC}   & -2.5866 & True & 3,931 \\
4  & \texttt{CCCCCC}   & -2.5866 & True & 3,887 \\
5  & \texttt{CCCCCC}   & -2.5866 & True & 4,138 \\
6  & \texttt{CCC=CCC}  & -2.3626 & True & 4,212 \\
7  & \texttt{CCCCCC}   & -2.5866 & True & 4,524 \\
8  & \texttt{CCCCCC}   & -2.5866 & True & 4,175 \\
9  & \texttt{CCCCCC}   & -2.5866 & True & 4,253 \\
10 & \texttt{CCCCCC}   & -2.5866 & True & 4,486 \\
11 & \texttt{CCCCC=C}  & -2.3626 & True & 4,446 \\
12 & \texttt{CCCCCC}   & -2.5866 & True & 4,038 \\
13 & \texttt{CCCCCC}   & -2.5866 & True & 4,273 \\
14 & \texttt{CCCCCC}   & -2.5866 & True & 4,048 \\
15 & \texttt{CCCCCC}   & -2.5866 & True & 3,830 \\
16 & \texttt{CCCCCC}   & -2.5866 & True & 4,508 \\
17 & \texttt{CCCCCC}   & -2.5866 & True & 4,066 \\
18 & \texttt{CCCCCC}   & -2.5866 & True & 4,237 \\
19 & \texttt{CCCCCC}   & -2.5866 & True & 4,141 \\
20 & \texttt{CCCCCC}   & -2.5866 & True & 4,048 \\
21 & \texttt{CCCCCC}   & -2.5866 & True & 4,568 \\
22 & \texttt{CCCCCC}   & -2.5866 & True & 4,417 \\
23 & \texttt{CCCCCC}   & -2.5866 & True & 3,983 \\
24 & \texttt{CCCCCC}   & -2.5866 & True & 3,951 \\
25 & \texttt{CCCCCC}   & -2.5866 & True & 3,798 \\
26 & \texttt{CCCCCC}   & -2.5866 & True & 3,898 \\
27 & \texttt{CCCCCC}   & -2.5866 & True & 3,973 \\
28 & \texttt{CCCCCC}   & -2.5866 & True & 4,047 \\
29 & \texttt{CCCCCC}   & -2.5866 & True & 3,812 \\
30 & \texttt{CCCCCC}   & -2.5866 & True & 4,348 \\
\bottomrule
\multicolumn{5}{l}{$^1$plogP.} \\
\end{tabular}
\end{table}

\begin{table}[ht!]\centering
\caption{Details on the QEVO simulations using 6 SELFIES tokens, random initialization and the Bologna-Yale ansatz. The number of the repetition, the SMILES string of the best state generated, its loss value, $\mathcal{L}$, if QEVO generates at least one state belonging to the top-10 loss value positions as characterized in the reference solution space, and the number of unique states generated during the current simulation are reported.}%
\begin{tabular}{@{}lllll@{}}
\toprule
Repetition & Best state           & $\mathcal{L}^1$  & In the top-10? & Unique states sampled \\ 
\midrule
1  & \texttt{FC=CCCC}    & -2.2697 & True  & 1,830 \\
2  & \texttt{CC=CCCC}    & -2.3626 & True  & 1,851 \\
3  & \texttt{CC=CCCC}    & -2.3626 & True  & 2,258 \\
4  & \texttt{CCCCCC}     & -2.5866 & True  & 1,501 \\
5  & \texttt{CCCC=CF}    & -2.2697 & True  & 1,399 \\
6  & \texttt{CCCCCC}     & -2.5866 & True  & 1,218 \\
7  & \texttt{CCCCCC}     & -2.5866 & True  & 1,677 \\
8  & \texttt{CC=CC=CC}   & -2.1386 & True  & 1,584 \\
9  & \texttt{C=CCC=CF}   & -2.0457 & True  & 1,814 \\
10 & \texttt{C=CCC=CF}   & -2.0457 & True  & 1,491 \\
11 & \texttt{CCCCCC}     & -2.5866 & True  & 2,439 \\
12 & \texttt{C=CCCCC}    & -2.3626 & True  & 1,431 \\
13 & \texttt{CCCCCC}     & -2.5866 & True  & 1,381 \\
14 & \texttt{CCCCCC}     & -2.5866 & True  & 1,880 \\
15 & \texttt{CCCCCC}     & -2.5866 & True  & 1,785 \\
16 & \texttt{FC=CCC=C}   & -2.0457 & True  & 2,239 \\
17 & \texttt{CCCCCC}     & -2.5866 & True  & 1,779 \\
18 & \texttt{CCCCCC}     & -2.5866 & True  & 1,447 \\
19 & \texttt{CCCCCC}     & -2.5866 & True  & 2,025 \\
20 & \texttt{C=CCCCC}    & -2.3626 & True  & 1,981 \\
21 & \texttt{CCCCCC}     & -2.5866 & True  & 1,499 \\
22 & \texttt{CCCCCC}     & -2.5866 & True  & 1,350 \\
23 & \texttt{CCCCCC}     & -2.5866 & True  & 1,932 \\
24 & \texttt{CCCCC=C}    & -2.3626 & True  & 1,295 \\
25 & \texttt{CCCCCC}     & -2.5866 & True  & 1,025 \\
26 & \texttt{CCCCCC}     & -2.5866 & True  & 1,184 \\
27 & \texttt{C=CCC=CC}   & -2.1386 & True  & 1,517 \\
28 & \texttt{C=CCCCC}    & -2.3626 & True  & 1,049 \\
29 & \texttt{CCCCCC}     & -2.5866 & True  & 1,694 \\
30 & \texttt{CC=CN=NF}   & -1.8567 & False & 1,094 \\
\bottomrule
\multicolumn{5}{l}{$^1$plogP.} \\
\end{tabular}
\end{table}

\begin{table}[ht!]\centering
\caption{Details on the QEVO simulations using 6 SELFIES tokens, random initialization and the RealAmplitudes ansatz. The number of the repetition, the SMILES string of the best state generated, its loss value, $\mathcal{L}$, if QEVO generates at least one state belonging to the top-10 loss value positions as characterized in the reference solution space, and the number of unique states generated during the current simulation are reported.}%
\begin{tabular}{@{}lllll@{}}
\toprule
Repetition & Best state           & $\mathcal{L}^1$  & In the top-10? & Unique states sampled \\ 
\midrule
1  & \texttt{CCCCCC}     & -2.5866 & True  & 1,768 \\
2  & \texttt{CCCCCC}     & -2.5866 & True  & 1,792 \\
3  & \texttt{CCCCCC}     & -2.5866 & True  & 2,156 \\
4  & \texttt{CCCCCC}     & -2.5866 & True  & 1,264 \\
5  & \texttt{FC=CC=CF}   & -1.9528 & False  & 2,018 \\
6  & \texttt{CCCCCC}     & -2.5866 & True  & 2,673 \\
7  & \texttt{CCCCCC}     & -2.5866 & True  & 1,992 \\
8  & \texttt{CCCC=CF}    & -2.2697 & True  & 1,884 \\
9  & \texttt{CCC=CCC}    & -2.3626 & True  & 1,521 \\
10 & \texttt{CCCCCC}     & -2.5866 & True  & 2,013 \\
11 & \texttt{CCCCCC}     & -2.5866 & True  & 2,319 \\
12 & \texttt{CCCCCC}     & -2.5866 & True  & 2,644 \\
13 & \texttt{CCCCCC}     & -2.5866 & True  & 1,228 \\
14 & \texttt{CCCCCC}     & -2.5866 & True  & 2,095 \\
15 & \texttt{CCCCCC}     & -2.5866 & True  & 2,663 \\
16 & \texttt{CC=CCCC}    & -2.3626 & True  & 2,278 \\
17 & \texttt{CCCCCC}     & -2.5866 & True  & 1,513 \\
18 & \texttt{CCCCCC}     & -2.5866 & True  & 2,019 \\
19 & \texttt{CCCCC}      & -2.1965 & True  & 1,522 \\
20 & \texttt{CCCCCC}     & -2.5866 & True  & 1,202 \\
21 & \texttt{CCCCCC}     & -2.5866 & True  & 2,188 \\
22 & \texttt{CCCCCC}     & -2.5866 & True  & 1,692 \\
23 & \texttt{CCC=CF}     & -1.8796 & False & 1,452 \\
24 & \texttt{CCCCCC}     & -2.5866 & True  & 2,215 \\
25 & \texttt{CCCCCC}     & -2.5866 & True  & 2,708 \\
26 & \texttt{CCCCCC}     & -2.5866 & True  & 1,542 \\
27 & \texttt{CCCCCC}     & -2.5866 & True  & 2,366 \\
28 & \texttt{C=CC=CCC}   & -2.1386 & True  & 1,698 \\
29 & \texttt{CCCCCC}     & -2.5866 & True  & 2,184 \\
30 & \texttt{CCCCCC}     & -2.5866 & True  & 2,065 \\
\bottomrule
\multicolumn{5}{l}{$^1$plogP.} \\
\end{tabular}
\end{table}

\begin{table}[ht!]\centering
\caption{Details on the QEVO simulations using 7 SELFIES tokens, uniform initialization and the Bologna-Yale ansatz. The number of the repetition, the SMILES string of the best state generated, its loss value, $\mathcal{L}$, if QEVO generates at least one state belonging to the top-10 loss value positions as characterized in the reference solution space, and the number of unique states generated during the current simulation are reported.}%
\begin{tabular}{@{}lllll@{}}
\toprule
Repetition & Best state           & $\mathcal{L}^1$  & In the top-10? & Unique states sampled \\
\midrule
1  & \texttt{CCCCCCC}     & -2.9767 & True  & 8,363 \\
2  & \texttt{CCC=CCCC}    & -2.7527 & True  & 7,777 \\
3  & \texttt{CCCCCCC}     & -2.9767 & True  & 8,628 \\
4  & \texttt{CCCCCCC}     & -2.9767 & True  & 9,045 \\
5  & \texttt{CCCCCCC}     & -2.9767 & True  & 8,002 \\
6  & \texttt{CCCCCCC}     & -2.9767 & True  & 8,302 \\
7  & \texttt{CCCCCCC}     & -2.9767 & True  & 8,089 \\
8  & \texttt{CCCCCCC}     & -2.9767 & True  & 8,156 \\
9  & \texttt{CCCCCCC}     & -2.9767 & True  & 8,115 \\
10 & \texttt{CCCCCCC}     & -2.9767 & True  & 8,123 \\
11 & \texttt{CCCCCCC}     & -2.9767 & True  & 8,987 \\
12 & \texttt{CCCCCCC}     & -2.9767 & True  & 7,669 \\
13 & \texttt{CCCCCCC}     & -2.9767 & True  & 8,399 \\
14 & \texttt{CC=CCC=CC}   & -2.5287 & True  & 7,776 \\
15 & \texttt{CCCCCCC}     & -2.9767 & True  & 8,549 \\
16 & \texttt{CCCCCCC}     & -2.9767 & True  & 7,988 \\
17 & \texttt{CCCCCCC}     & -2.9767 & True  & 8,353 \\
18 & \texttt{CC=CCCCC}    & -2.7527 & True  & 8,119 \\
19 & \texttt{CCCCCCC}     & -2.9767 & True  & 7,923 \\
20 & \texttt{CCCCCCC}     & -2.9767 & True  & 10,047 \\
21 & \texttt{CCCCCCC}     & -2.9767 & True  & 8,463 \\
22 & \texttt{CCCCCC=C}    & -2.7527 & True  & 8,538 \\
23 & \texttt{CCCCC=CF}    & -2.6598 & True  & 8,614 \\
24 & \texttt{CCCCCCC}     & -2.9767 & True  & 9,382 \\
25 & \texttt{CCCCCCC}     & -2.9767 & True  & 8,081 \\
26 & \texttt{C=CCCC=CC}   & -2.5287 & True  & 8,857 \\
27 & \texttt{CCCCCCC}     & -2.9767 & True  & 8,375 \\
28 & \texttt{CCCCCCC}     & -2.9767 & True  & 8,443 \\
29 & \texttt{CCCCCCC}     & -2.9767 & True  & 7,981 \\
30 & \texttt{CCCCCCC}     & -2.9767 & True  & 8,104 \\
\bottomrule
\multicolumn{5}{l}{$^1$plogP.} \\
\end{tabular}
\end{table}

\begin{table}[ht!]\centering
\caption{Details on the QEVO simulations using 7 SELFIES tokens, uniform initialization and the RealAmplitudes ansatz. The number of the repetition, the SMILES string of the best state generated, its loss value, $\mathcal{L}$, if QEVO generates at least one state belonging to the top-10 loss value positions as characterized in the reference solution space, and the number of unique states generated during the current simulation are reported.}%
\begin{tabular}{@{}lllll@{}}
\toprule
Repetition & Best state           & $\mathcal{L}^1$  & In the top-10? & Unique states sampled \\
\midrule
1  & \texttt{CCCCCCC}     & -2.9767 & True  & 8,547 \\
2  & \texttt{CCCCCCC}     & -2.9767 & True  & 7,285 \\
3  & \texttt{CCCCCCC}     & -2.9767 & True  & 9,419 \\
4  & \texttt{C=CC=CCCC}   & -2.5287 & True  & 7,150 \\
5  & \texttt{CCCCCCC}     & -2.9767 & True  & 8,738 \\
6  & \texttt{CCCCCCC}     & -2.9767 & True  & 9,290 \\
7  & \texttt{CCCCCCC}     & -2.9767 & True  & 8,054 \\
8  & \texttt{CCCCCCC}     & -2.9767 & True  & 7,684 \\
9  & \texttt{CCCCCCC}     & -2.9767 & True  & 8,163 \\
10 & \texttt{CCCCCCC}     & -2.9767 & True  & 7,500 \\
11 & \texttt{CCCCCCC}     & -2.9767 & True  & 7,859 \\
12 & \texttt{CCCCCCC}     & -2.9767 & True  & 8,353 \\
13 & \texttt{CCCCCCC}     & -2.9767 & True  & 8,446 \\
14 & \texttt{CCCCCCC}     & -2.9767 & True  & 8,257 \\
15 & \texttt{CCCCCCC}     & -2.9767 & True  & 8,325 \\
16 & \texttt{CC=CCCCC}    & -2.7527 & True  & 8,100 \\
17 & \texttt{CCCCCCC}     & -2.9767 & True  & 7,946 \\
18 & \texttt{CCCCC=CC}    & -2.7527 & True  & 7,899 \\
19 & \texttt{CCCCCCC}     & -2.9767 & True  & 9,435 \\
20 & \texttt{CCCCC=CC}    & -2.7527 & True  & 9,232 \\
21 & \texttt{CCCCCCC}     & -2.9767 & True  & 7,773 \\
22 & \texttt{CCCCCCC}     & -2.9767 & True  & 7,491 \\
23 & \texttt{CCCCCCC}     & -2.9767 & True  & 9,524 \\
24 & \texttt{CCCCCCC}     & -2.9767 & True  & 8,522 \\
25 & \texttt{CCCCCCC}     & -2.9767 & True  & 8,460 \\
26 & \texttt{CCCCCCC}     & -2.9767 & True  & 8,015 \\
27 & \texttt{CCCCCCC}     & -2.9767 & True  & 8,551 \\
28 & \texttt{CCCCCCC}     & -2.9767 & True  & 9,570 \\
29 & \texttt{CCCCCCC}     & -2.9767 & True  & 7,730 \\
30 & \texttt{CCCCCCC}     & -2.9767 & True  & 7,366 \\
\bottomrule
\multicolumn{5}{l}{$^1$plogP.} \\
\end{tabular}
\end{table}

\begin{table}[ht!]\centering
\caption{Details on the QEVO simulations using 7 SELFIES tokens, random initialization and the Bologna-Yale ansatz. The number of the repetition, the SMILES string of the best state generated, its loss value, $\mathcal{L}$, if QEVO generates at least one state belonging to the top-10 loss value positions as characterized in the reference solution space, and the number of unique states generated during the current simulation are reported.}%
\begin{tabular}{@{}lllll@{}}
\toprule
Repetition & Best state           & $\mathcal{L}^1$  & In the top-10? & Unique states sampled \\
\midrule
1  & \texttt{C=CC=C=CCC}  & -2.2937 & False & 3,067 \\
2  & \texttt{FC=CCC}      & -1.8796 & False & 2,044 \\
3  & \texttt{CCCCCCC}     & -2.9767 & True  & 4,580 \\
4  & \texttt{CCC=CCCC}    & -2.7527 & True  & 2,434 \\
5  & \texttt{CCCCCCC}     & -2.9767 & True  & 1,966 \\
6  & \texttt{C=CCCC=CF}   & -2.4358 & True  & 3,191 \\
7  & \texttt{CCCCCCC}     & -2.9767 & True  & 3,230 \\
8  & \texttt{CCCCCCC}     & -2.9767 & True  & 2,775 \\
9  & \texttt{CCCCC=CF}    & -2.6598 & True  & 2,867 \\
10 & \texttt{CC=CCCCC}    & -2.7527 & True  & 4,301 \\
11 & \texttt{CCCC=CF}     & -2.2697 & False & 2,668 \\
12 & \texttt{CCCCCCC}     & -2.9767 & True  & 4,125 \\
13 & \texttt{CCC=C=CCC}   & -2.5177 & True  & 2,427 \\
14 & \texttt{CCCCCC=C}    & -2.7527 & True  & 3,792 \\
15 & \texttt{CC=CCC=CC}   & -2.5287 & True  & 3,319 \\
16 & \texttt{CCCCCC}      & -2.5866 & True  & 2,648 \\
17 & \texttt{CCCC=CC=C}   & -2.5287 & True  & 3,099 \\
18 & \texttt{CCCCCCC}     & -2.9767 & True  & 2,992 \\
19 & \texttt{CCCCCC}      & -2.5866 & True  & 1,987 \\
20 & \texttt{CCCCCCC}     & -2.9767 & True  & 3,133 \\
21 & \texttt{CC=CC=CCC}   & -2.5287 & True  & 2,720 \\
22 & \texttt{CCCC=CN=N}   & -2.3312 & False & 2,999 \\
23 & \texttt{CCCCCCC}     & -2.9767 & True  & 4,312 \\
24 & \texttt{CCCCCCC}     & -2.9767 & True  & 3,106 \\
25 & \texttt{CCCCCCC}     & -2.9767 & True  & 2,659 \\
26 & \texttt{CCC=CCCC}    & -2.7527 & True  & 3,367 \\
27 & \texttt{CCCCCCC}     & -2.9767 & True  & 3,420 \\
28 & \texttt{CCCCCCC}     & -2.9767 & True  & 4,312 \\
29 & \texttt{CCCC=CC=C}   & -2.5287 & True  & 1,987 \\
30 & \texttt{CCCCCCC}     & -2.9767 & True  & 3,918 \\
\bottomrule
\multicolumn{5}{l}{$^1$plogP.} \\
\end{tabular}
\end{table}

\begin{table}[ht!]\centering
\caption{Details on the QEVO simulations using 7 SELFIES tokens, random initialization and the RealAmplitudes ansatz. The number of the repetition, the SMILES string of the best state generated, its loss value, $\mathcal{L}$, if QEVO generates at least one state belonging to the top-10 loss value positions as characterized in the reference solution space, and the number of unique states generated during the current simulation are reported.}%
\begin{tabular}{@{}lllll@{}}
\toprule
Repetition & Best state           & $\mathcal{L}^1$  & In the top-10? & Unique states sampled \\
\midrule
1  & \texttt{CCCCCCC}     & -2.9767 & True  & 4,988 \\
2  & \texttt{CCCCCCC}     & -2.9767 & True  & 2,984 \\
3  & \texttt{CCCCCCC}     & -2.9767 & True  & 5,015 \\
4  & \texttt{CCCCCCC}     & -2.9767 & True  & 2,948 \\
5  & \texttt{CCCCCCC}     & -2.9767 & True  & 4,524 \\
6  & \texttt{CCCCC=CC}    & -2.7527 & True  & 3,826 \\
7  & \texttt{CC=CCCCC}    & -2.7527 & True  & 2,960 \\
8  & \texttt{CCCCCCC}     & -2.9767 & True  & 3,154 \\
9  & \texttt{C=CCCCCC}    & -2.7527 & True  & 1,968 \\
10 & \texttt{FN=NC=CF}    & -1.7638 & False & 967  \\
11 & \texttt{CCCCC}       & -2.1965 & False & 3,636 \\
12 & \texttt{CCCCCCC}     & -2.9767 & True  & 2,728 \\
13 & \texttt{CCCCCCC}     & -2.9767 & True  & 2,486 \\
14 & \texttt{CCCCCCC}     & -2.9767 & True  & 4,066 \\
15 & \texttt{CCCCCCC}     & -2.9767 & True  & 2,583 \\
16 & \texttt{CCCCCCC}     & -2.9767 & True  & 2,536 \\
17 & \texttt{CCCCCCC}     & -2.9767 & True  & 5,178 \\
18 & \texttt{CCCCCCC}     & -2.9767 & True  & 3,130 \\
19 & \texttt{CCCCCCC}     & -2.9767 & True  & 5,975 \\
20 & \texttt{CC=CCCCC}    & -2.7527 & True  & 3,568 \\
21 & \texttt{CCCCCCC}     & -2.9767 & True  & 4,746 \\
22 & \texttt{CCCCCCC}     & -2.9767 & True  & 6,385 \\
23 & \texttt{CCCCCCC}     & -2.9767 & True  & 4,376 \\
24 & \texttt{C=CCCCCC}    & -2.7527 & True  & 3,522 \\
25 & \texttt{CCCC=CCC}    & -2.7527 & True  & 4,560 \\
26 & \texttt{CCCCCCC}     & -2.9767 & True  & 4,350 \\
27 & \texttt{CCCCCCC}     & -2.9767 & True  & 4,039 \\
28 & \texttt{CCCCCCC}     & -2.9767 & True  & 1,936 \\
29 & \texttt{CCCCCCC}     & -2.9767 & True  & 6,240 \\
30 & \texttt{CCCCCCC}     & -2.9767 & True  & 3,622 \\
\bottomrule
\multicolumn{5}{l}{$^1$plogP.} \\
\end{tabular}
\end{table}

\begin{table}[ht!]\centering
\caption{Details on the QEVO simulations using 8 SELFIES tokens, uniform initialization and the Bologna-Yale ansatz. The number of the repetition, the SMILES string of the best state generated, its loss value, $\mathcal{L}$, if QEVO generates at least one state belonging to the top-10 loss value positions as characterized in the reference solution space, and the number of unique states generated during the current simulation are reported.}%
\begin{tabular}{@{}lllll@{}}
\toprule
Repetition & Best state           & $\mathcal{L}^1$  & In the top-10? & Unique states sampled \\ 
\midrule
1  & \texttt{CC=CCCCC}     & -2.7527 & True  & 14,645 \\
2  & \texttt{CCCCCCCC}     & -3.3668 & True  & 16,141 \\
3  & \texttt{CCCCCCCC}     & -3.3668 & True  & 14,694 \\
4  & \texttt{CCCCC=CF}     & -2.6598 & False & 12,283 \\
5  & \texttt{CCC=CCCCC}    & -3.1428 & True  & 15,346 \\
6  & \texttt{CCCCCCCC}     & -3.3668 & True  & 15,136 \\
7  & \texttt{CCC=CCCC}     & -2.7527 & True  & 14,243 \\
8  & \texttt{CCCCC=CF}     & -3.0499 & True  & 12,375 \\
9  & \texttt{CCCCCCC=C}    & -3.1428 & True  & 13,765 \\
10 & \texttt{CCCCCCCC}     & -3.3668 & True  & 16,796 \\
11 & \texttt{CC=CCCCCF}    & -2.7023 & False & 12,241 \\
12 & \texttt{C=CCCCCCC}    & -3.1428 & True  & 13,154 \\
13 & \texttt{C=CCCCCC=C}   & -2.9188 & True  & 13,341 \\
14 & \texttt{CCCCCCCC}     & -3.3668 & True  & 16,110 \\
15 & \texttt{CCCCCC}       & -2.5866 & False & 16,003 \\
16 & \texttt{CCCCCCCC}     & -3.3668 & True  & 15,121 \\
17 & \texttt{CCCC=C=CC=C}  & -2.6838 & False & 15,127 \\
18 & \texttt{CCCCCC}       & -2.5866 & False & 12,230 \\
19 & \texttt{CCCCCCC}      & -2.9767 & True  & 17,246 \\
20 & \texttt{OC=CCCCCC}    & -2.6384 & False & 12,386 \\
21 & \texttt{CCCCCCCC}     & -3.3668 & True  & 15,133 \\
22 & \texttt{CCCCCCCC}     & -3.3668 & True  & 16,599 \\
23 & \texttt{CCCCCCCC}     & -3.3668 & True  & 16,846 \\
24 & \texttt{CCCCCCCC}     & -3.3668 & True  & 15,957 \\
25 & \texttt{CCCCCCC=C}    & -3.1428 & True  & 15,260 \\
26 & \texttt{CCCCCCCC}     & -3.3668 & True  & 16,035 \\
27 & \texttt{CCCCCCCC}     & -3.3668 & True  & 17,033 \\
28 & \texttt{CCCC=CCCC}    & -3.1428 & True  & 15,051 \\
29 & \texttt{CCCCCC}       & -2.5866 & False & 13,059 \\
30 & \texttt{CCCCCCCC}     & -3.3668 & True  & 14,555 \\
\bottomrule
\multicolumn{5}{l}{$^1$plogP.} \\
\end{tabular}
\end{table}

\begin{table}[ht!]\centering
\caption{Details on the QEVO simulations using 8 SELFIES tokens, uniform initialization and the RealAmplitudes ansatz. The number of the repetition, the SMILES string of the best state generated, its loss value, $\mathcal{L}$, if QEVO generates at least one state belonging to the top-10 loss value positions as characterized in the reference solution space, and the number of unique states generated during the current simulation are reported.}%
\begin{tabular}{@{}lllll@{}}
\toprule
Repetition & Best state           & $\mathcal{L}^1$  & In the top-10? & Unique states sampled \\ 
\midrule
1  & \texttt{CCCCCCCC}     & -3.3668 & True  & 14,943 \\
2  & \texttt{CCC=C=CCCC}   & -2.9078 & True  & 14,892 \\
3  & \texttt{CC=CCCCCC}    & -3.1428 & True  & 18,079 \\
4  & \texttt{CCCC=CCCC}    & -3.1428 & True  & 17,573 \\
5  & \texttt{CCCCCCCC}     & -3.3668 & True  & 18,546 \\
6  & \texttt{CCCCCCCC}     & -3.3668 & True  & 15,336 \\
7  & \texttt{CCC=CCCCC}    & -3.1428 & True  & 15,812 \\
8  & \texttt{CC=CC=CCCC}   & -2.9188 & True  & 11,754 \\
9  & \texttt{CCC=CC=CCC}   & -2.9188 & True  & 17,134 \\
10 & \texttt{CCCCCCCC}     & -3.3668 & True  & 13,806 \\
11 & \texttt{CCCCCCCC}     & -3.3668 & True  & 15,042 \\
12 & \texttt{CCCCCCCC}     & -3.3668 & True  & 13,569 \\
13 & \texttt{CCCC=CCCC}    & -3.1428 & True  & 13,139 \\
14 & \texttt{CCCC=CCCC}    & -3.1428 & True  & 15,310 \\
15 & \texttt{CCCCCCCC}     & -3.3668 & True  & 15,922 \\
16 & \texttt{CCCCCCCC}     & -3.3668 & True  & 13,163 \\
17 & \texttt{CCCCCCCC}     & -3.3668 & True  & 14,796 \\
18 & \texttt{CCCCCCCC}     & -3.3668 & True  & 15,560 \\
19 & \texttt{CCCCCCCC}     & -3.3668 & True  & 20,242 \\
20 & \texttt{CCCCCCCC}     & -3.3668 & True  & 19,717 \\
21 & \texttt{CCCCCCCC}     & -3.3668 & True  & 20,434 \\
22 & \texttt{CCCCCCCC}     & -3.3668 & True  & 14,742 \\
23 & \texttt{CCCCCCCC}     & -3.3668 & True  & 17,196 \\
24 & \texttt{CCCCCCCC}     & -3.3668 & True  & 14,770 \\
25 & \texttt{FC=CC=CCCC}   & -2.8259 & True  & 15,054 \\
26 & \texttt{CCCCCCCC}     & -3.3668 & True  & 19,496 \\
27 & \texttt{CCCCCCCC}     & -3.3668 & True  & 18,382 \\
28 & \texttt{CC=CCCCCC}    & -3.1428 & True  & 21,311 \\
29 & \texttt{CCCCCCCC}     & -3.3668 & True  & 14,445 \\
30 & \texttt{CCCCCCCC}     & -3.3668 & True  & 16,026 \\
\bottomrule
\multicolumn{5}{l}{$^1$plogP.} \\
\end{tabular}
\end{table}

\begin{table}[ht!]\centering
\caption{Details on the QEVO simulations using 8 SELFIES tokens, random initialization and the Bologna-Yale ansatz. The number of the repetition, the SMILES string of the best state generated, its loss value, $\mathcal{L}$, if QEVO generates at least one state belonging to the top-10 loss value positions as characterized in the reference solution space, and the number of unique states generated during the current simulation are reported.}%
\begin{tabular}{@{}lllll@{}}
\toprule
Repetition & Best state           & $\mathcal{L}^1$  & In the top-10? & Unique states sampled \\
\midrule
1  & \texttt{CCCCCCCC}     & -3.3668 & True  & 4,850 \\
2  & \texttt{CCCCCCC=CF}   & -3.0499 & True  & 6,156 \\
3  & \texttt{CCCCCCCC}     & -3.3668 & True  & 4,656 \\
4  & \texttt{FC=CCCCCC}    & -3.0499 & True  & 4,686 \\
5  & \texttt{CCCCCCCC}     & -3.3668 & True  & 6,154 \\
6  & \texttt{CCCCCCCC}     & -3.3668 & True  & 6,562 \\
7  & \texttt{CCCCCCC}      & -2.5866 & False & 4,970 \\
8  & \texttt{FC=CCCCCC}    & -3.0499 & True  & 6,741 \\
9  & \texttt{CCCCCCC=CF}   & -3.0499 & True  & 6,278 \\
10 & \texttt{C=C=C=CCCCC}  & -2.6728 & False & 2,160 \\
11 & \texttt{CCC=CC=C=CF}  & -2.5909 & False & 3,869 \\
12 & \texttt{CCCCCCCC}     & -3.3668 & True  & 4,600 \\
13 & \texttt{CCCCCCCC}     & -3.3668 & True  & 5,093 \\
14 & \texttt{CCCCCCCCF}    & -2.9263 & True  & 4,856 \\
15 & \texttt{CCCCCCCC}     & -3.3668 & True  & 5,543 \\
16 & \texttt{FCCC=C=CCC}   & -2.4673 & False & 4,727 \\
17 & \texttt{C=C=C=CCC=CC} & -2.4488 & False & 6,498 \\
18 & \texttt{CCCCCCCC}     & -3.3668 & True  & 5,558 \\
19 & \texttt{CCCCCCC=CC}   & -3.1428 & True  & 5,878 \\
20 & \texttt{CCC=C=CCCC}   & -2.9078 & True  & 8,309 \\
21 & \texttt{CCCCCCCC}     & -3.3668 & True  & 7,563 \\
22 & \texttt{CCCCCCCC}     & -3.3668 & True  & 6,332 \\
23 & \texttt{CCCCCCCC}     & -3.3668 & True  & 7,405 \\
24 & \texttt{CCCCCCCC}     & -3.3668 & True  & 4,400 \\
25 & \texttt{CCCCCCCC}     & -3.3668 & True  & 4,194 \\
26 & \texttt{CCCCCCC=CC}   & -3.1428 & True  & 7,040 \\
27 & \texttt{CCCCCCCC}     & -3.3668 & True  & 5,459 \\
28 & \texttt{CCC=CC=CCC}   & -2.9188 & True  & 4,986 \\
29 & \texttt{CCC=CCCC=C}   & -2.9188 & True  & 4,386 \\
30 & \texttt{CCCCC=CF}     & -2.6598 & False & 4,718 \\
\bottomrule
\multicolumn{5}{l}{$^1$plogP.} \\
\end{tabular}
\end{table}

\begin{table}[ht!]\centering
\caption{Details on the QEVO simulations using 8 SELFIES tokens, random initialization and the RealAmplitudes ansatz. The number of the repetition, the SMILES string of the best state generated, its loss value, $\mathcal{L}$, if QEVO generates at least one state belonging to the top-10 loss value positions as characterized in the reference solution space, and the number of unique states generated during the current simulation are reported.}%
\begin{tabular}{@{}lllll@{}}
\toprule
Repetition & Best state           & $\mathcal{L}^1$  & In the top-10? & Unique states sampled \\
\midrule
1  & \texttt{CCCCCCCC}     & -3.3668 & True  & 6,844 \\
2  & \texttt{CCCCCCCC}     & -3.3668 & True  & 6,799 \\
3  & \texttt{CCCCCCC=C}    & -3.1428 & True  & 7,339 \\
4  & \texttt{CCCCCCCC}     & -3.3668 & True  & 5,428 \\
5  & \texttt{CCC=CCC=CF}   & -2.8259 & True  & 6,943 \\
6  & \texttt{CCCCCCCC}     & -3.3668 & True  & 9,073 \\
7  & \texttt{CCCCCCCC}     & -3.3668 & True  & 10,500 \\
8  & \texttt{FC=CCCCCC}    & -3.0499 & True  & 5,161 \\
9  & \texttt{CCCCCCCC}     & -3.3668 & True  & 8,581 \\
10 & \texttt{CCCCCCCC}     & -3.3668 & True  & 4,670 \\
11 & \texttt{CCCCCCCC}     & -3.3668 & True  & 7,494 \\
12 & \texttt{CCCCCCCC}     & -3.3668 & True  & 6,089 \\
13 & \texttt{CCCCCCCC}     & -3.3668 & True  & 5,784 \\
14 & \texttt{CC=CCC=C=CC}  & -2.6838 & False & 5,911 \\
15 & \texttt{CCCCCCCC}     & -3.3668 & True  & 5,129 \\
16 & \texttt{CCCCCCCC}     & -3.3668 & True  & 6,241 \\
17 & \texttt{CCCCCCCC}     & -3.3668 & True  & 7,067 \\
18 & \texttt{CCCCCCCC}     & -3.3668 & True  & 5,148 \\
19 & \texttt{FCCCCCCCC}    & -2.9263 & True  & 5,874 \\
20 & \texttt{CCCCCCCC}     & -3.3668 & True  & 9,153 \\
21 & \texttt{FN=NC=CCCC}   & -2.6369 & False & 4,326 \\
22 & \texttt{CCCCCCCC}     & -3.3668 & True  & 4,378 \\
23 & \texttt{CCCCCCCC}     & -3.3668 & True  & 6,507 \\
24 & \texttt{CCCCCCCC}     & -3.3668 & True  & 7,806 \\
25 & \texttt{CCCCCCC=CC}   & -3.1428 & True  & 11,232 \\
26 & \texttt{CCCCCCCC}     & -3.3668 & True  & 6,241 \\
27 & \texttt{CCCCCCCC}     & -3.3668 & True  & 7,287 \\
28 & \texttt{CCCCCCCC}     & -3.3668 & True  & 7,834 \\
29 & \texttt{CCCCCCCC}     & -3.3668 & True  & 11,316 \\
30 & \texttt{CCCCCCC=C}    & -3.1428 & True  & 8,858 \\
\bottomrule
\multicolumn{5}{l}{$^1$plogP.} \\
\end{tabular}
\end{table}

\begin{table}[ht!]\centering
\caption{Details on the QEVO simulations using 9 SELFIES tokens, uniform initialization and the Bologna-Yale ansatz. The number of the repetition, the SMILES string of the best state generated, its loss value, $\mathcal{L}$, if QEVO generates at least one state belonging to the top-10 loss value positions as characterized in the reference solution space, and the number of unique states generated during the current simulation are reported.}%
\begin{tabular}{@{}lllll@{}}
\toprule
Repetition & Best state           & $\mathcal{L}^1$    & In the top-10? & Unique states sampled \\ 
\midrule
1  & \texttt{CCC=C=C=CCC=C}    & -2.8389 & False & 21,695 \\
2  & \texttt{CCCC=C=CCC=C}     & -3.0739 & False & 24,951 \\
3  & \texttt{FC=CCC=CCC}       & -2.8259 & False & 22,956 \\
4  & \texttt{CCC=CCC=CF}       & -2.8259 & False & 22,122 \\
5  & \texttt{CCCCCCC=CF}       & -3.4400 & True  & 22,711 \\
6  & \texttt{CCCCCCCCC}        & -3.7569 & True  & 22,104 \\
7  & \texttt{CCCCCCC=CF}       & -3.4400 & True  & 21,586 \\
\end{tabular}
\end{table}

\begin{table}[ht!]\centering
\caption{Details on the QEVO simulations using 9 SELFIES tokens, uniform initialization and the RealAmplitudes ansatz. The number of the repetition, the SMILES string of the best state generated, its loss value, $\mathcal{L}$, if QEVO generates at least one state belonging to the top-10 loss value positions as characterized in the reference solution space, and the number of unique states generated during the current simulation are reported.}%
\begin{tabular}{@{}lllll@{}}
\toprule
Repetition & Best state           & $\mathcal{L}^1$    & In the top-10? & Unique states sampled \\ 
\midrule
1  & \texttt{CCCCCCC=CF}        & -3.4400 & True  & 19,861 \\
2  & \texttt{CCCCCCCCC}         & -3.7569 & True  & 23,288 \\
3  & \texttt{C=CCCCCCCC}        & -3.5329 & True  & 24,204 \\
4  & \texttt{CCCCCCCCC}         & -3.7569 & True  & 22,577 \\
5  & \texttt{CC=CC=CCCC=C}      & -3.0849 & False & 21,520 \\
6  & \texttt{CCCCCCCCC}         & -3.7569 & True  & 23,071 \\
7  & \texttt{CCCCCCCCC}         & -3.7569 & True  & 29,735 \\
\end{tabular}
\end{table}

\begin{table}[ht!]\centering
\caption{Details on the QEVO simulations using 9 SELFIES tokens, random initialization and the Bologna-Yale ansatz. The number of the repetition, the SMILES string of the best state generated, its loss value, $\mathcal{L}$, if QEVO generates at least one state belonging to the top-10 loss value positions as characterized in the reference solution space, and the number of unique states generated during the current simulation are reported.}%
\begin{tabular}{@{}lllll@{}}
\toprule
Repetition & Best state           & $\mathcal{L}^1$    & In the top-10? & Unique states sampled \\ 
\midrule
1  & \texttt{CCCCCCC=CF}        & -3.4400 & True  & 9,317  \\
2  & \texttt{FC=CCCCCCC}        & -3.4400 & True  & 6,301  \\
3  & \texttt{CCCCCCCC}          & -3.3668 & True  & 7,059  \\
4  & \texttt{CCCCCCCCC}         & -3.7569 & True  & 12,717 \\
5  & \texttt{CCCCCCCC=C}        & -3.5329 & True  & 10,657 \\
6  & \texttt{CCCCCCCCC}         & -3.7569 & True  & 8,176  \\
7  & \texttt{CCCCCCCCC}         & -3.7569 & True  & 8,248  \\
\end{tabular}
\end{table}

\begin{table}[ht!]\centering
\caption{Details on the QEVO simulations using 9 SELFIES tokens, random initialization and the RealAmplitudes ansatz. The number of the repetition, the SMILES string of the best state generated, its loss value, $\mathcal{L}$, if QEVO generates at least one state belonging to the top-10 loss value positions as characterized in the reference solution space, and the number of unique states generated during the current simulation are reported.}%
\begin{tabular}{@{}lllll@{}}
\toprule
Repetition & Best state           & $\mathcal{L}^1$    & In the top-10? & Unique states sampled \\ 
\midrule
1  & \texttt{CCCCCCCCC}         & -3.7569 & True  & 10,672 \\
2  & \texttt{CCCCCCCCC}         & -3.7569 & True  & 12,359 \\
3  & \texttt{CC=CCC=C=CCF}      & -2.6334 & False & 4,420  \\
4  & \texttt{CCCCCCCCC}         & -3.7569 & True  & 11,832 \\
5  & \texttt{CCCCCCCCC}         & -3.7569 & True  & 13,562 \\
6  & \texttt{CCCC=CCCCC}        & -3.5329 & True  & 10,637 \\
7  & \texttt{CCCCCCCCC}         & -3.7569 & True  & 8,907  \\
\end{tabular}
\end{table}

\begin{table}[ht!]\centering
\caption{Details on the QEVO simulations using 6 SELFIES tokens, uniform initialization and the Bologna-Yale ansatz. The number of the repetition, the SMILES string of the best state generated, its loss value, $\mathcal{L}$, if QEVO generates at least one state belonging to the top-10 loss value positions as characterized in the reference solution space, and the number of unique states generated during the current simulation are reported.}
\begin{tabular}{@{}lllll@{}}
\toprule
Repetition & Best state           & $\mathcal{L}^1$    & In the top-10? & Unique states sampled \\ 
\midrule
1  & \texttt{CCCCCO} & 0.3767 & True & 4,917 \\
2  & \texttt{CCCCCO} & 0.3767 & True & 5,041 \\
3  & \texttt{CCCCCO} & 0.3767 & True & 5,422 \\
4  & \texttt{CCCCCO} & 0.3767 & True & 4,795 \\
5  & \texttt{CCCCCO} & 0.3767 & True & 4,532 \\
6  & \texttt{OCCCCC} & 0.3767 & True & 5,228 \\
7  & \texttt{OCCCCC} & 0.3767 & True & 4,863 \\
8  & \texttt{OCCCCC} & 0.3767 & True & 4,731 \\
9  & \texttt{CCCCCO} & 0.3767 & True & 4,744 \\
10 & \texttt{CCCCCO} & 0.3767 & True & 5,185 \\
11 & \texttt{OCCCCC} & 0.3767 & True & 5,550 \\
12 & \texttt{OCCCCC} & 0.3767 & True & 5,127 \\
13 & \texttt{CCCCCO} & 0.3767 & True & 4,934 \\
14 & \texttt{OCCCCC} & 0.3767 & True & 4,800 \\
15 & \texttt{CCCCCO} & 0.3767 & True & 5,128 \\
16 & \texttt{CCCCCO} & 0.3767 & True & 5,276 \\
17 & \texttt{CCCCCO} & 0.3767 & True & 5,055 \\
18 & \texttt{CCCCCO} & 0.3767 & True & 4,868 \\
19 & \texttt{CCCCCO} & 0.3767 & True & 5,116 \\
20 & \texttt{CCCCCO} & 0.3767 & True & 4,922 \\
21 & \texttt{OCCCCC} & 0.3767 & True & 5,335 \\
22 & \texttt{CCCCCO} & 0.3767 & True & 5,250 \\
23 & \texttt{OCCCCC} & 0.3767 & True & 5,112 \\
24 & \texttt{CCCCCO} & 0.3767 & True & 5,100 \\
25 & \texttt{CCCCCO} & 0.3767 & True & 4,683 \\
26 & \texttt{CCCCCO} & 0.3767 & True & 5,055 \\
27 & \texttt{OCCCCC} & 0.3767 & True & 4,671 \\
28 & \texttt{CCCCCO} & 0.3767 & True & 5,454 \\
29 & \texttt{OCCCCC} & 0.3767 & True & 5,245 \\
30 & \texttt{OCCCCC} & 0.3767 & True & 4,465 \\
\bottomrule
\multicolumn{5}{l}{$^1$Unbiased drug design.} \\
\end{tabular}
\end{table}

\begin{table}[ht!]\centering
\caption{Details on the QEVO simulations using 6 SELFIES tokens, uniform initialization and the RealAmplitudes ansatz. The number of the repetition, the SMILES string of the best state generated, its loss value, $\mathcal{L}$, if QEVO generates at least one state belonging to the top-10 loss value positions as characterized in the reference solution space, and the number of unique states generated during the current simulation are reported.}
\begin{tabular}{@{}lllll@{}}
\toprule
Repetition & Best state           & $\mathcal{L}^1$    & In the top-10? & Unique states sampled \\ 
\midrule
1  & \texttt{CCCCCO} & 0.3767 & True & 4,488 \\
2  & \texttt{OCCCCC} & 0.3767 & True & 4,618 \\
3  & \texttt{CCCCCO} & 0.3767 & True & 4,485 \\
4  & \texttt{CCCCCO} & 0.3767 & True & 4,540 \\
5  & \texttt{CCCCCO} & 0.3767 & True & 4,527 \\
6  & \texttt{OCCCCC} & 0.3767 & True & 4,955 \\
7  & \texttt{CCCCCO} & 0.3767 & True & 4,606 \\
8  & \texttt{CCCCCO} & 0.3767 & True & 4,405 \\
9  & \texttt{OCCCCC} & 0.3767 & True & 4,868 \\
10 & \texttt{OCCCCC} & 0.3767 & True & 4,545 \\
11 & \texttt{OCCCCC} & 0.3767 & True & 4,944 \\
12 & \texttt{CCCCCO} & 0.3767 & True & 4,547 \\
13 & \texttt{OCCCCC} & 0.3767 & True & 4,524 \\
14 & \texttt{CCCCCO} & 0.3767 & True & 4,571 \\
15 & \texttt{CCCCCO} & 0.3767 & True & 5,037 \\
16 & \texttt{OCCCCC} & 0.3767 & True & 4,696 \\
17 & \texttt{CCCCCO} & 0.3767 & True & 4,837 \\
18 & \texttt{CCCCCO} & 0.3767 & True & 4,953 \\
19 & \texttt{CCCCCO} & 0.3767 & True & 4,620 \\
20 & \texttt{CCCCCO} & 0.3767 & True & 4,585 \\
21 & \texttt{OCCCCC} & 0.3767 & True & 5,183 \\
22 & \texttt{CCCCCO} & 0.3767 & True & 5,166 \\
23 & \texttt{OCCCCC} & 0.3767 & True & 4,463 \\
24 & \texttt{CCCCCO} & 0.3767 & True & 4,456 \\
25 & \texttt{CCCCCO} & 0.3767 & True & 4,795 \\
26 & \texttt{CCCCCO} & 0.3767 & True & 4,577 \\
27 & \texttt{CCCCCO} & 0.3767 & True & 4,947 \\
28 & \texttt{OCCCCC} & 0.3767 & True & 4,892 \\
29 & \texttt{CCCCCO} & 0.3767 & True & 4,506 \\
30 & \texttt{CCCCCO} & 0.3767 & True & 4,832 \\
\bottomrule
\multicolumn{5}{l}{$^1$Unbiased drug design.} \\
\end{tabular}
\end{table}

\begin{table}[ht!]\centering
\caption{Details on the QEVO simulations using 6 SELFIES tokens, random initialization and the Bologna-Yale ansatz. The number of the repetition, the SMILES string of the best state generated, its loss value, $\mathcal{L}$, if QEVO generates at least one state belonging to the top-10 loss value positions as characterized in the reference solution space, and the number of unique states generated during the current simulation are reported.}
\begin{tabular}{@{}lllll@{}}
\toprule
Repetition & Best state           & $\mathcal{L}^1$    & In the top-10? & Unique states sampled \\ 
\midrule
1  & \texttt{OCCCCC} & 0.3767 & True & 1,875 \\
2  & \texttt{OCCCC}  & 0.3874 & True & 1,438 \\
3  & \texttt{OCCCCC} & 0.3767 & True & 1,695 \\
4  & \texttt{OCCCCC} & 0.3767 & True & 1,697 \\
5  & \texttt{OCCCCC} & 0.3767 & True & 2,184 \\
6  & \texttt{OCCCC}  & 0.3874 & True & 1,347 \\
7  & \texttt{NCCCCC} & 0.3783 & True & 1,967 \\
8  & \texttt{CCCCCO} & 0.3767 & True & 2,695 \\
9  & \texttt{CCCCN}  & 0.3783 & True & 2,122 \\
10 & \texttt{OCCCCC} & 0.3767 & True & 2,206 \\
11 & \texttt{CCCCCO} & 0.3767 & True & 2,112 \\
12 & \texttt{CCCCO}  & 0.3874 & True & 1,509 \\
13 & \texttt{OCCCC}  & 0.3874 & True & 1,134 \\
14 & \texttt{CCCCO}  & 0.3767 & True & 1,886 \\
15 & \texttt{CCCCN}  & 0.3783 & True & 2,181 \\
16 & \texttt{OCCCC}  & 0.3874 & True & 2,238 \\
17 & \texttt{OCCCCC} & 0.3767 & True & 2,053 \\
18 & \texttt{OCCCCC} & 0.3767 & True & 2,055 \\
19 & \texttt{OCCCCC} & 0.3767 & True & 2,615 \\
20 & \texttt{CCCCO}  & 0.3767 & True & 2,479 \\
21 & \texttt{CCCCO}  & 0.3767 & True & 1,676 \\
22 & \texttt{CCCCO}  & 0.3767 & True & 1,503 \\
23 & \texttt{OCCCCC} & 0.3767 & True & 1,880 \\
24 & \texttt{OCCCCC} & 0.3767 & True & 1,563 \\
25 & \texttt{CCCCN}  & 0.3909 & True & 1,221 \\
26 & \texttt{NCCCC}  & 0.3909 & True & 1,381 \\
27 & \texttt{CCCCO}  & 0.3767 & True & 2,013 \\
28 & \texttt{CCCCO}  & 0.3767 & True & 1,508 \\
29 & \texttt{CCCNC}  & 0.3893 & True & 1,915 \\
30 & \texttt{OCCCC}  & 0.3874 & True & 1,553 \\
\bottomrule
\multicolumn{5}{l}{$^1$Unbiased drug design.} \\
\end{tabular}
\end{table}

\begin{table}[ht!]\centering
\caption{Details on the QEVO simulations using 6 SELFIES tokens, random initialization and the RealAmplitudes ansatz. The number of the repetition, the SMILES string of the best state generated, its loss value, $\mathcal{L}$, if QEVO generates at least one state belonging to the top-10 loss value positions as characterized in the reference solution space, and the number of unique states generated during the current simulation are reported.}
\begin{tabular}{@{}lllll@{}}
\toprule
Repetition & Best state           & $\mathcal{L}^1$    & In the top-10? & Unique states sampled \\ 
\midrule
1  & \texttt{OCCCCC} & 0.3767 & True & 1,786 \\
2  & \texttt{CCCCO}  & 0.3767 & True & 1,949 \\
3  & \texttt{CCCCO}  & 0.3767 & True & 2,518 \\
4  & \texttt{CCCCO}  & 0.3767 & True & 2,137 \\
5  & \texttt{CCCCO}  & 0.3767 & True & 2,728 \\
6  & \texttt{CCCCO}  & 0.3767 & True & 2,281 \\
7  & \texttt{CCCCO}  & 0.3767 & True & 2,303 \\
8  & \texttt{CCCCO}  & 0.3767 & True & 1,671 \\
9  & \texttt{CCCCO}  & 0.3767 & True & 1,820 \\
10 & \texttt{CCCCO}  & 0.3767 & True & 1,542 \\
11 & \texttt{OCCCCC} & 0.3767 & True & 2,531 \\
12 & \texttt{CCCCO}  & 0.3767 & True & 2,189 \\
13 & \texttt{OCCCCC} & 0.3767 & True & 1,428 \\
14 & \texttt{CCCCO}  & 0.3767 & True & 2,944 \\
15 & \texttt{OCCCCC} & 0.3767 & True & 2,815 \\
16 & \texttt{CCCCO}  & 0.3767 & True & 2,642 \\
17 & \texttt{OCCCCC} & 0.3767 & True & 1,986 \\
18 & \texttt{CCCCO}  & 0.3767 & True & 2,342 \\
19 & \texttt{CCCCO}  & 0.3767 & True & 2,363 \\
20 & \texttt{CCCCO}  & 0.3767 & True & 2,170 \\
21 & \texttt{CCCCO}  & 0.3767 & True & 2,443 \\
22 & \texttt{OCCCCC} & 0.3767 & True & 2,150 \\
23 & \texttt{CCCCO}  & 0.3767 & True & 3,125 \\
24 & \texttt{CCCCO}  & 0.3767 & True & 2,351 \\
25 & \texttt{CCCCO}  & 0.3767 & True & 1,722 \\
26 & \texttt{NCCCCC} & 0.3783 & True & 1,635 \\
27 & \texttt{CCCCO}  & 0.3767 & True & 2,562 \\
28 & \texttt{CCCCO}  & 0.3767 & True & 2,275 \\
29 & \texttt{OCCCCC} & 0.3767 & True & 1,682 \\
30 & \texttt{OCCCCC} & 0.3767 & True & 2,765 \\
\bottomrule
\multicolumn{5}{l}{$^1$Unbiased drug design.} \\
\end{tabular}
\end{table}

\begin{table}[ht!]\centering
\caption{Details on the QEVO simulations using 7 SELFIES tokens, uniform initialization and the Bologna-Yale ansatz. The number of the repetition, the SMILES string of the best state generated, its loss value, $\mathcal{L}$, if QEVO generates at least one state belonging to the top-10 loss value positions as characterized in the reference solution space, and the number of unique states generated during the current simulation are reported.}
\begin{tabular}{@{}lllll@{}}
\toprule
Repetition & Best state           & $\mathcal{L}^1$    & In the top-10? & Unique states sampled \\ 
\midrule
1  & \texttt{CCCCCCO}  & 0.3628 & True & 16,545 \\
2  & \texttt{CCCCCCO}  & 0.3628 & True & 14,594 \\
3  & \texttt{CCCCCCO}  & 0.3628 & True & 17,198 \\
4  & \texttt{CCCCCCO}  & 0.3628 & True & 12,211 \\
5  & \texttt{CCCCCCN}  & 0.3641 & True & 12,036 \\
6  & \texttt{CCCCCCO}  & 0.3628 & True & 12,612 \\
7  & \texttt{CCCCCCO}  & 0.3628 & True & 14,928 \\
8  & \texttt{CCCCCCO}  & 0.3628 & True & 11,928 \\
9  & \texttt{CNCCCCC}  & 0.3727 & True & 11,404 \\
10 & \texttt{CCCCCNC}  & 0.3727 & True & 10,993 \\
11 & \texttt{CCCCCCO}  & 0.3628 & True & 13,475 \\
12 & \texttt{CCCCCCO}  & 0.3628 & True & 13,014 \\
13 & \texttt{CCCNCCC}  & 0.3720 & True & 12,287 \\
14 & \texttt{CCCCCCO}  & 0.3628 & True & 11,713 \\
15 & \texttt{OCCCCCC}  & 0.3628 & True & 12,025 \\
16 & \texttt{CCCCCCO}  & 0.3628 & True & 17,946 \\
17 & \texttt{CNCCCCC}  & 0.3727 & True & 11,942 \\
18 & \texttt{OCCCCCC}  & 0.3628 & True & 10,703 \\
19 & \texttt{CCCCCCO}  & 0.3628 & True & 12,256 \\
20 & \texttt{OCCCCCC}  & 0.3628 & True & 14,370 \\
21 & \texttt{CNCCCCC}  & 0.3727 & True & 12,230 \\
22 & \texttt{CCCCCCO}  & 0.3628 & True & 12,168 \\
23 & \texttt{OCCCCCC}  & 0.3628 & True & 10,439 \\
24 & \texttt{CCCCCNC}  & 0.3727 & True & 13,337 \\
25 & \texttt{CCCCCCO}  & 0.3628 & True & 13,041 \\
26 & \texttt{CCCCCCO}  & 0.3628 & True & 10,693 \\
27 & \texttt{CCCNCCC}  & 0.3720 & True & 12,172 \\
28 & \texttt{CCCCCCO}  & 0.3628 & True & 10,939 \\
29 & \texttt{CCCCCCO}  & 0.3628 & True & 12,538 \\
30 & \texttt{CCCCCCN}  & 0.3641 & True & 12,085 \\
\bottomrule
\multicolumn{5}{l}{$^1$Unbiased drug design.} \\
\end{tabular}
\end{table}

\begin{table}[ht!]\centering
\caption{Details on the QEVO simulations using 7 SELFIES tokens, uniform initialization and the RealAmplitudes ansatz. The number of the repetition, the SMILES string of the best state generated, its loss value, $\mathcal{L}$, if QEVO generates at least one state belonging to the top-10 loss value positions as characterized in the reference solution space, and the number of unique states generated during the current simulation are reported.}
\begin{tabular}{@{}lllll@{}}
\toprule
Repetition & Best state           & $\mathcal{L}^1$    & In the top-10? & Unique states sampled \\ 
\midrule
1  & \texttt{CCCCCCO}  & 0.3628 & True & 10,801 \\
2  & \texttt{OCCCCCC}  & 0.3628 & True & 11,392 \\
3  & \texttt{CCCCCCO}  & 0.3628 & True & 10,715 \\
4  & \texttt{CCCCCCO}  & 0.3628 & True & 10,705 \\
5  & \texttt{CCCCCCO}  & 0.3628 & True & 10,512 \\
6  & \texttt{CCCCCCO}  & 0.3628 & True & 11,976 \\
7  & \texttt{CCCCCCO}  & 0.3628 & True & 10,801 \\
8  & \texttt{OCCCCCC}  & 0.3628 & True & 10,394 \\
9  & \texttt{CCCCCCO}  & 0.3628 & True & 10,732 \\
10 & \texttt{CCCCCCO}  & 0.3628 & True & 13,068 \\
11 & \texttt{CCCCCCO}  & 0.3628 & True & 11,078 \\
12 & \texttt{CCCCCCO}  & 0.3628 & True & 10,916 \\
13 & \texttt{CCCCCCO}  & 0.3628 & True & 11,261 \\
14 & \texttt{CCCCCCO}  & 0.3628 & True & 13,952 \\
15 & \texttt{OCCCCCC}  & 0.3628 & True & 14,074 \\
16 & \texttt{OCCCCCC}  & 0.3628 & True & 11,434 \\
17 & \texttt{OCCCCCC}  & 0.3628 & True & 11,346 \\
18 & \texttt{CCCCCCO}  & 0.3628 & True & 10,979 \\
19 & \texttt{CCCCCCO}  & 0.3628 & True & 11,528 \\
20 & \texttt{CCCCCCO}  & 0.3628 & True & 11,688 \\
21 & \texttt{CCCCCCO}  & 0.3628 & True & 10,751 \\
22 & \texttt{CCCCCCO}  & 0.3628 & True & 10,593 \\
23 & \texttt{CCCCCCO}  & 0.3628 & True & 12,129 \\
24 & \texttt{CCCCCCO}  & 0.3628 & True & 10,546 \\
25 & \texttt{OCCCCCC}  & 0.3628 & True & 10,323 \\
26 & \texttt{CCCCCCO}  & 0.3628 & True & 12,065 \\
27 & \texttt{OCCCCCC}  & 0.3628 & True & 11,243 \\
28 & \texttt{CCCNCCC}  & 0.3720 & True & 10,046 \\
29 & \texttt{CCCCCCO}  & 0.3628 & True & 11,171 \\
30 & \texttt{CCCCCCO}  & 0.3628 & True & 11,622 \\
\bottomrule
\multicolumn{5}{l}{$^1$Unbiased drug design.} \\
\end{tabular}
\end{table}

\begin{table}[ht!]\centering
\caption{Details on the QEVO simulations using 7 SELFIES tokens, random initialization and the Bologna-Yale ansatz. The number of the repetition, the SMILES string of the best state generated, its loss value, $\mathcal{L}$, if QEVO generates at least one state belonging to the top-10 loss value positions as characterized in the reference solution space, and the number of unique states generated during the current simulation are reported.}
\begin{tabular}{@{}lllll@{}}
\toprule
Repetition & Best state           & $\mathcal{L}^1$    & In the top-10? & Unique states sampled \\ 
\midrule
1  & \texttt{OCCCCCC}   & 0.3628 & True & 3,362 \\
2  & \texttt{CCCCCCO}   & 0.3628 & True & 3,999 \\
3  & \texttt{OCCCCCC}   & 0.3628 & True & 5,036 \\
4  & \texttt{OCCCCCC}   & 0.3628 & True & 3,390 \\
5  & \texttt{OCCCCOC}   & 0.3769 & True & 2,227 \\
6  & \texttt{CCCCCCO}   & 0.3628 & True & 4,732 \\
7  & \texttt{CCCCCCO}   & 0.3628 & True & 4,090 \\
8  & \texttt{OCCCCCC}   & 0.3628 & True & 5,065 \\
9  & \texttt{OCCCCCC}   & 0.3628 & True & 5,761 \\
10 & \texttt{CCNCCCC}   & 0.3735 & True & 2,815 \\
11 & \texttt{CCCCCCN}   & 0.3641 & True & 3,921 \\
12 & \texttt{OCCCCC}    & 0.3767 & True & 3,372 \\
13 & \texttt{CCCCCCN}   & 0.3641 & True & 3,005 \\
14 & \texttt{OCCCCC}    & 0.3767 & True & 3,277 \\
15 & \texttt{OCCCCCC}   & 0.3628 & True & 4,174 \\
16 & \texttt{CCCCCCN}   & 0.3641 & True & 2,791 \\
17 & \texttt{OCCCCCC}   & 0.3628 & True & 3,510 \\
18 & \texttt{CCCCCCN}   & 0.3641 & True & 2,951 \\
19 & \texttt{OCCCCC}    & 0.3767 & True & 3,370 \\
20 & \texttt{OCCCC}     & 0.3874 & False & 2,358 \\
21 & \texttt{CCCCO}     & 0.3874 & False & 4,484 \\
22 & \texttt{CCCCCCO}   & 0.3628 & True & 3,376 \\
23 & \texttt{OCCCOCC}   & 0.3819 & True & 2,649 \\
24 & \texttt{CCCCCCO}   & 0.3628 & True & 3,457 \\
25 & \texttt{CCCCO}     & 0.3874 & False & 3,268 \\
26 & \texttt{NCCCCOC}   & 0.3791 & True & 2,775 \\
27 & \texttt{OCCCCCC}   & 0.3628 & True & 4,732 \\
28 & \texttt{CCCNCCC}   & 0.3720 & True & 3,417 \\
29 & \texttt{OCCCCCC}   & 0.3628 & True & 2,436 \\
30 & \texttt{CCNCCCC}   & 0.3735 & True & 4,200 \\
\bottomrule
\multicolumn{5}{l}{$^1$Unbiased drug design.} \\
\end{tabular}
\end{table}

\begin{table}[ht!]\centering
\caption{Details on the QEVO simulations using 7 SELFIES tokens, random initialization and the RealAmplitudes ansatz. The number of the repetition, the SMILES string of the best state generated, its loss value, $\mathcal{L}$, if QEVO generates at least one state belonging to the top-10 loss value positions as characterized in the reference solution space, and the number of unique states generated during the current simulation are reported.}
\begin{tabular}{@{}lllll@{}}
\toprule
Repetition & Best state           & $\mathcal{L}^1$    & In the top-10? & Unique states sampled \\ 
\midrule
1  & \texttt{OCCCC}     & 0.3874 & False & 4,599 \\
2  & \texttt{OCCCCCC}   & 0.3628 & True  & 3,650 \\
3  & \texttt{OCCCCCC}   & 0.3628 & True  & 4,170 \\
4  & \texttt{CCCCCCO}   & 0.3628 & True  & 5,389 \\
5  & \texttt{CCCCCCO}   & 0.3628 & True  & 7,270 \\
6  & \texttt{CCCCCCO}   & 0.3628 & True  & 4,454 \\
7  & \texttt{CCCCNCC}   & 0.3735 & True  & 4,332 \\
8  & \texttt{CCCCCCO}   & 0.3628 & True  & 3,683 \\
9  & \texttt{OCCCCCC}   & 0.3628 & True  & 2,810 \\
10 & \texttt{OCCCCOC}   & 0.3769 & True  & 2,707 \\
11 & \texttt{CCCCCCN}   & 0.3641 & True  & 4,548 \\
12 & \texttt{OCCCCCC}   & 0.3628 & True  & 2,645 \\
13 & \texttt{OCCCCCC}   & 0.3628 & True  & 3,579 \\
14 & \texttt{CCOCCCO}   & 0.3819 & True  & 3,558 \\
15 & \texttt{OCCCCCC}   & 0.3628 & True  & 3,454 \\
16 & \texttt{CCCCCCO}   & 0.3628 & True  & 3,746 \\
17 & \texttt{CCCCCCO}   & 0.3628 & True  & 6,508 \\
18 & \texttt{CNCCCCC}   & 0.3727 & True  & 2,299 \\
19 & \texttt{CCCCCCO}   & 0.3628 & True  & 3,821 \\
20 & \texttt{NCCCCCC}   & 0.3641 & True  & 5,803 \\
21 & \texttt{CCCCCCO}   & 0.3628 & True  & 4,371 \\
22 & \texttt{CCCCCCO}   & 0.3628 & True  & 4,767 \\
23 & \texttt{OCCCC}     & 0.3874 & False & 4,436 \\
24 & \texttt{CCCCCCN}   & 0.3641 & True  & 5,587 \\
25 & \texttt{CCCCO}     & 0.3767 & True  & 4,532 \\
26 & \texttt{OCCCCCC}   & 0.3628 & True  & 3,402 \\
27 & \texttt{OCCCCCC}   & 0.3628 & True  & 2,636 \\
28 & \texttt{OCCCCC}    & 0.3767 & True  & 2,745 \\
29 & \texttt{OCCCCCC}   & 0.3628 & True  & 6,456 \\
30 & \texttt{C(O)CCC}   & 0.3874 & False & 2,747 \\
\bottomrule
\multicolumn{5}{l}{$^1$Unbiased drug design.} \\
\end{tabular}
\end{table}

\begin{table}[ht!]\centering
\caption{Details on the QEVO simulations using 8 SELFIES tokens, uniform initialization and the Bologna-Yale ansatz. The number of the repetition, the SMILES string of the best state generated, its loss value, $\mathcal{L}$, if QEVO generates at least one state belonging to the top-10 loss value positions as characterized in the reference solution space, and the number of unique states generated during the current simulation are reported.}
\begin{tabular}{@{}lllll@{}}
\toprule
Repetition & Best state           & $\mathcal{L}^1$    & In the top-10? & Unique states sampled \\ 
\midrule
1  & \texttt{CCCCCCCO}   & 0.3560 & True  & 22,553 \\
2  & \texttt{CCCCO}      & 0.3767 & False & 21,667 \\
3  & \texttt{CCCCCCO}    & 0.3628 & True  & 22,015 \\
4  & \texttt{C(CO)CCC}   & 0.3767 & False & 20,218 \\
5  & \texttt{OCCCCCC}    & 0.3628 & True  & 26,659 \\
6  & \texttt{CCCCCCN}    & 0.3641 & True  & 20,177 \\
7  & \texttt{CCCCNCC}    & 0.3646 & True  & 27,849 \\
8  & \texttt{NCCCCOCC}   & 0.3702 & False & 18,703 \\
9  & \texttt{CNCCCCCC}   & 0.3648 & True  & 18,723 \\
10 & \texttt{CCCCCCO}    & 0.3628 & True  & 21,434 \\
11 & \texttt{NCCCCCC}    & 0.3641 & True  & 28,811 \\
12 & \texttt{NCCCCCC}    & 0.3641 & True  & 20,417 \\
13 & \texttt{CCCCCCCN}   & 0.3571 & True  & 19,845 \\
14 & \texttt{CCNCCCC}    & 0.3735 & False & 24,545 \\
15 & \texttt{CCCCCCN}    & 0.3641 & True  & 23,690 \\
16 & \texttt{CCCCCCO}    & 0.3628 & True  & 27,989 \\
17 & \texttt{O=C(O)CCC}  & 0.3625 & True  & 21,654 \\
18 & \texttt{CCCCCNC}    & 0.3727 & False & 19,034 \\
19 & \texttt{CCCNCCC}    & 0.3720 & False & 16,360 \\
20 & \texttt{CCCCCCO}    & 0.3628 & True  & 19,882 \\
21 & \texttt{CCCCCCCO}   & 0.3560 & True  & 42,736 \\
22 & \texttt{CCCCNCC}    & 0.3646 & True  & 22,374 \\
23 & \texttt{O=C(O)CCC}  & 0.3625 & True  & 26,268 \\
24 & \texttt{COCCCCCN}   & 0.3662 & True  & 22,505 \\
25 & \texttt{OCCCCCC}    & 0.3628 & True  & 50,961 \\
26 & \texttt{CCCCO}      & 0.3767 & False & 20,819 \\
27 & \texttt{CCCCO}      & 0.3767 & False & 17,914 \\
28 & \texttt{CCCCCCCO}   & 0.3560 & True  & 18,067 \\
29 & \texttt{O=C(O)CCC}  & 0.3625 & True  & 16,980 \\
30 & \texttt{COCCCCNC}   & 0.3743 & False & 22,857 \\
\bottomrule
\multicolumn{5}{l}{$^1$Unbiased drug design.} \\
\end{tabular}
\end{table}

\begin{table}[ht!]\centering
\caption{Details on the QEVO simulations using 8 SELFIES tokens, uniform initialization and the RealAmplitudes ansatz. The number of the repetition, the SMILES string of the best state generated, its loss value, $\mathcal{L}$, if QEVO generates at least one state belonging to the top-10 loss value positions as characterized in the reference solution space, and the number of unique states generated during the current simulation are reported.}
\begin{tabular}{@{}lllll@{}}
\toprule
Repetition & Best state           & $\mathcal{L}^1$    & In the top-10? & Unique states sampled \\ 
\midrule
1  & \texttt{CCCCCCCO}   & 0.3560 & True  & 22,918 \\
2  & \texttt{OCCCCCC}    & 0.3628 & True  & 19,954 \\
3  & \texttt{CCCCCCCO}   & 0.3560 & True  & 20,986 \\
4  & \texttt{CCCCCCCO}   & 0.3560 & True  & 22,441 \\
5  & \texttt{OCCOCCCC}   & 0.3682 & False & 19,749 \\
6  & \texttt{CCCCCCCO}   & 0.3560 & True  & 21,597 \\
7  & \texttt{OCCCCCCC}   & 0.3560 & True  & 20,698 \\
8  & \texttt{CCCCCCCO}   & 0.3560 & True  & 19,740 \\
9  & \texttt{OCCCCCCC}   & 0.3560 & True  & 20,845 \\
10 & \texttt{CCCCCCCO}   & 0.3560 & True  & 20,126 \\
11 & \texttt{CCCCCCCO}   & 0.3560 & True  & 20,465 \\
12 & \texttt{CCCCCCCO}   & 0.3560 & True  & 19,888 \\
13 & \texttt{CCCCCCCO}   & 0.3560 & True  & 22,814 \\
14 & \texttt{OCCCCCCC}   & 0.3560 & True  & 19,411 \\
15 & \texttt{NCCCCCCC}   & 0.3571 & True  & 19,193 \\
16 & \texttt{OCCCCCCC}   & 0.3560 & True  & 20,144 \\
17 & \texttt{CCCCCCCO}   & 0.3560 & True  & 21,740 \\
18 & \texttt{CCCCCCCO}   & 0.3560 & True  & 19,960 \\
19 & \texttt{CCCNCCCC}   & 0.3640 & True  & 25,117 \\
20 & \texttt{CCCNCCCC}   & 0.3640 & True  & 21,019 \\
21 & \texttt{CCCCCCCO}   & 0.3560 & True  & 24,959 \\
22 & \texttt{CCCCCCCO}   & 0.3560 & True  & 18,210 \\
23 & \texttt{CCCCCCCO}   & 0.3560 & True  & 23,973 \\
24 & \texttt{CCCCCCCO}   & 0.3560 & True  & 20,879 \\
25 & \texttt{CCCCCCCO}   & 0.3560 & True  & 21,525 \\
26 & \texttt{OCCCCCC}    & 0.3628 & True  & 18,042 \\
27 & \texttt{OCCCCCCC}   & 0.3560 & True  & 21,278 \\
28 & \texttt{CCCCCCCO}   & 0.3560 & True  & 22,572 \\
29 & \texttt{CCCCCCCO}   & 0.3560 & True  & 21,338 \\
30 & \texttt{NCCCCCCC}   & 0.3571 & True  & 19,152 \\
\bottomrule
\multicolumn{5}{l}{$^1$Unbiased drug design.} \\
\end{tabular}
\end{table}

\begin{table}[ht!]\centering
\caption{Details on the QEVO simulations using 8 SELFIES tokens, random initialization and the Bologna-Yale ansatz. The number of the repetition, the SMILES string of the best state generated, its loss value, $\mathcal{L}$, if QEVO generates at least one state belonging to the top-10 loss value positions as characterized in the reference solution space, and the number of unique states generated during the current simulation are reported.}
\begin{tabular}{@{}lllll@{}}
\toprule
Repetition & Best state           & $\mathcal{L}^1$    & In the top-10? & Unique states sampled \\ 
\midrule
1  & \texttt{NCCCCCCC}   & 0.3571 & True  & 5,610 \\
2  & \texttt{O=C(O)CCC}  & 0.3625 & True  & 5,237 \\
3  & \texttt{CCCCCO}     & 0.3767 & False & 4,547 \\
4  & \texttt{CCCCNCC}    & 0.3646 & True  & 5,627 \\
5  & \texttt{CCCCCCCO}   & 0.3560 & True  & 7,030 \\
6  & \texttt{CCCCCCCO}   & 0.3560 & True  & 6,007 \\
7  & \texttt{OCCCCCCC}   & 0.3560 & True  & 6,596 \\
8  & \texttt{NCCCCCCC}   & 0.3571 & True  & 8,729 \\
9  & \texttt{OCCCCCCC}   & 0.3560 & True  & 6,747 \\
10 & \texttt{C(C)CCN}    & 0.3909 & False & 4,423 \\
11 & \texttt{CCCCCCO}    & 0.3628 & True  & 5,418 \\
12 & \texttt{CCCCCCCO}   & 0.3560 & True  & 4,492 \\
13 & \texttt{C(O)CCCC}   & 0.3767 & False & 4,463 \\
14 & \texttt{OCCCCCCC}   & 0.3560 & True  & 4,264 \\
15 & \texttt{NCCCCCCC}   & 0.3571 & True  & 5,570 \\
16 & \texttt{C(CCCO)C}   & 0.3767 & False & 6,009 \\
17 & \texttt{OCCCCCCC}   & 0.3560 & True  & 10,270 \\
18 & \texttt{NCCCCCOC}   & 0.3662 & True  & 6,097 \\
19 & \texttt{OCCCCCCC}   & 0.3560 & True  & 6,552 \\
20 & \texttt{OCCCCC}     & 0.3767 & False & 6,728 \\
21 & \texttt{CCCCCCCO}   & 0.3560 & True  & 9,388 \\
22 & \texttt{CCCCCCCN}   & 0.3571 & True  & 9,771 \\
23 & \texttt{CCCCCCCO}   & 0.3560 & True  & 8,931 \\
24 & \texttt{CCCCCCN}    & 0.3571 & True  & 6,349 \\
25 & \texttt{CCCCNCCC}   & 0.3640 & True  & 5,888 \\
26 & \texttt{OCCCCCCC}   & 0.3560 & True  & 5,458 \\
27 & \texttt{CCCCCO}     & 0.3767 & False & 5,923 \\
28 & \texttt{C(O)CCC}    & 0.3874 & False & 5,801 \\
29 & \texttt{OCCCCCC}    & 0.3628 & True  & 6,558 \\
30 & \texttt{CCCCNCC}    & 0.3646 & True  & 5,613 \\
\bottomrule
\multicolumn{5}{l}{$^1$Unbiased drug design.} \\
\end{tabular}
\end{table}

\begin{table}[ht!]\centering
\caption{Details on the QEVO simulations using 8 SELFIES tokens, random initialization and the RealAmplitudes ansatz. The number of the repetition, the SMILES string of the best state generated, its loss value, $\mathcal{L}$, if QEVO generates at least one state belonging to the top-10 loss value positions as characterized in the reference solution space, and the number of unique states generated during the current simulation are reported.}
\begin{tabular}{@{}lllll@{}}
\toprule
Repetition & Best state           & $\mathcal{L}^1$    & In the top-10? & Unique states sampled \\ 
\midrule
1  & \texttt{CCCCCCCO}   & 0.3560 & True  & 7,779 \\
2  & \texttt{CCCCCCCO}   & 0.3560 & True  & 7,727 \\
3  & \texttt{OCCCCC}     & 0.3767 & False & 5,526 \\
4  & \texttt{CCCCCO}     & 0.3767 & False & 6,282 \\
5  & \texttt{CCCCCCCO}   & 0.3560 & True  & 11,431 \\
6  & \texttt{CCCCNCC}    & 0.3735 & False & 8,097 \\
7  & \texttt{CCCCCCN}    & 0.3641 & True  & 5,783 \\
8  & \texttt{CCCCNCC}    & 0.3735 & False & 4,360 \\
9  & \texttt{CCCCCCCO}   & 0.3560 & True  & 10,758 \\
10 & \texttt{CCCCCCCO}   & 0.3560 & True  & 8,518 \\
11 & \texttt{OCCCCCCC}   & 0.3560 & True  & 8,680 \\
12 & \texttt{CCCCCCO}    & 0.3628 & True  & 5,838 \\
13 & \texttt{OCCCCC}     & 0.3767 & False & 6,120 \\
14 & \texttt{CCCCCCCO}   & 0.3560 & True  & 6,766 \\
15 & \texttt{CCCCCO}     & 0.3767 & False & 4,717 \\
16 & \texttt{CCCCCCCO}   & 0.3560 & True  & 7,037 \\
17 & \texttt{CCCCCO}     & 0.3767 & False & 5,953 \\
18 & \texttt{CCCOCCCO}   & 0.3718 & False & 7,010 \\
19 & \texttt{CCCCNCCC}   & 0.3640 & True  & 5,327 \\
20 & \texttt{COCCCCCO}   & 0.3643 & True  & 7,931 \\
21 & \texttt{C(O)CCCC}   & 0.3767 & False & 5,235 \\
22 & \texttt{CCOCCCCO}   & 0.3683 & False & 7,759 \\
23 & \texttt{CCCCCCCO}   & 0.3560 & True  & 7,563 \\
24 & \texttt{COCCNCCC}   & 0.3716 & False & 7,122 \\
25 & \texttt{CCNCCCCC}   & 0.3646 & True  & 8,086 \\
26 & \texttt{CCCCCO}     & 0.3767 & False & 3,952 \\
27 & \texttt{CCCCCCCO}   & 0.3560 & True  & 6,289 \\
28 & \texttt{OCCCCCC}    & 0.3628 & True  & 7,293 \\
29 & \texttt{CCCCCCCO}   & 0.3560 & True  & 7,156 \\
30 & \texttt{CCCCCCCO}   & 0.3560 & True  & 5,006 \\
\bottomrule
\multicolumn{5}{l}{$^1$Unbiased drug design.} \\
\end{tabular}
\end{table}

\begin{table}[ht!]\centering
\caption{Details on the QEVO simulations using 9 SELFIES tokens, uniform initialization and the Bologna-Yale ansatz. The number of the repetition, the SMILES string of the best state generated, its loss value, $\mathcal{L}$, if QEVO generates at least one state belonging to the top-10 loss value positions as characterized in the reference solution space, and the number of unique states generated during the current simulation are reported.}
\begin{tabular}{@{}lllll@{}}
\toprule
Repetition & Best state           & $\mathcal{L}^1$    & In the top-10? & Unique states sampled \\ 
\midrule
1  & \texttt{CCCNC(C)C}      & 0.3480 & True  & 32,538 \\
2  & \texttt{CC(C)NCCC}      & 0.3480 & True  & 33,710 \\
3  & \texttt{CC(C)CNCC}      & 0.3541 & True  & 25,877 \\
4  & \texttt{OCCCCCC}        & 0.3628 & False & 31,499 \\
5  & \texttt{C(C)NCCCC}      & 0.3735 & False & 31,521 \\
6  & \texttt{NCCCCCCCC}      & 0.3553 & True  & 34,959 \\
7  & \texttt{CCCNCCC}        & 0.3720 & False & 32,709 \\
8  & \texttt{O=C(N)CCCC}     & 0.3520 & True  & 72,784 \\
9  & \texttt{CC(C)NCCC}      & 0.3480 & True  & 34,779 \\
10 & \texttt{CCCCC(C)O}      & 0.3621 & False & 41,003 \\
11 & \texttt{CC(CCCC)O}      & 0.3621 & False & 23,705 \\
12 & \texttt{CCC(C)NCC}      & 0.3692 & False & 31,483 \\
13 & \texttt{CCCCCCN}        & 0.3641 & False & 41,400 \\
14 & \texttt{CCCCCCCCO}      & 0.3544 & True  & 58,701 \\
15 & \texttt{CC(C)NCCC}      & 0.3480 & True  & 42,429 \\
16 & \texttt{CC(O)CCCC}      & 0.3621 & False & 34,989 \\
17 & \texttt{NCCOCCCCC}      & 0.3651 & False & 33,041 \\
18 & \texttt{NCCCCCCC}       & 0.3571 & False & 47,907 \\
19 & \texttt{CCCCCCO}        & 0.3628 & False & 28,581 \\
20 & \texttt{CC(CCCO)C}      & 0.3485 & True  & 58,065 \\
21 & \texttt{CCCNC(C)C}      & 0.3480 & True  & 29,734 \\
22 & \texttt{CC(C)CCCN}      & 0.3494 & True  & 37,338 \\
23 & \texttt{OCCCCCC}        & 0.3628 & False & 26,055 \\
24 & \texttt{CCCCC(C)O}      & 0.3621 & False & 38,524 \\
25 & \texttt{CC(CCNC)C}      & 0.3567 & False & 31,015 \\
26 & \texttt{OCCCCCC}        & 0.3628 & False & 39,253 \\
27 & \texttt{CCCCC(O)C}      & 0.3621 & False & 33,888 \\
28 & \texttt{CC(C)CCCO}      & 0.3485 & True  & 40,120 \\
29 & \texttt{NC(C)CCCC}      & 0.3639 & False & 39,243 \\
30 & \texttt{CCCNC(C)C}      & 0.3480 & True  & 38,864 \\
\bottomrule
\multicolumn{5}{l}{$^1$Unbiased drug design.} \\
\end{tabular}%
\end{table}

\clearpage
\pagebreak

\subsection{Drugs design}

\begin{figure}[!htbp]
    \centering
    \includegraphics[width=1.0\linewidth]{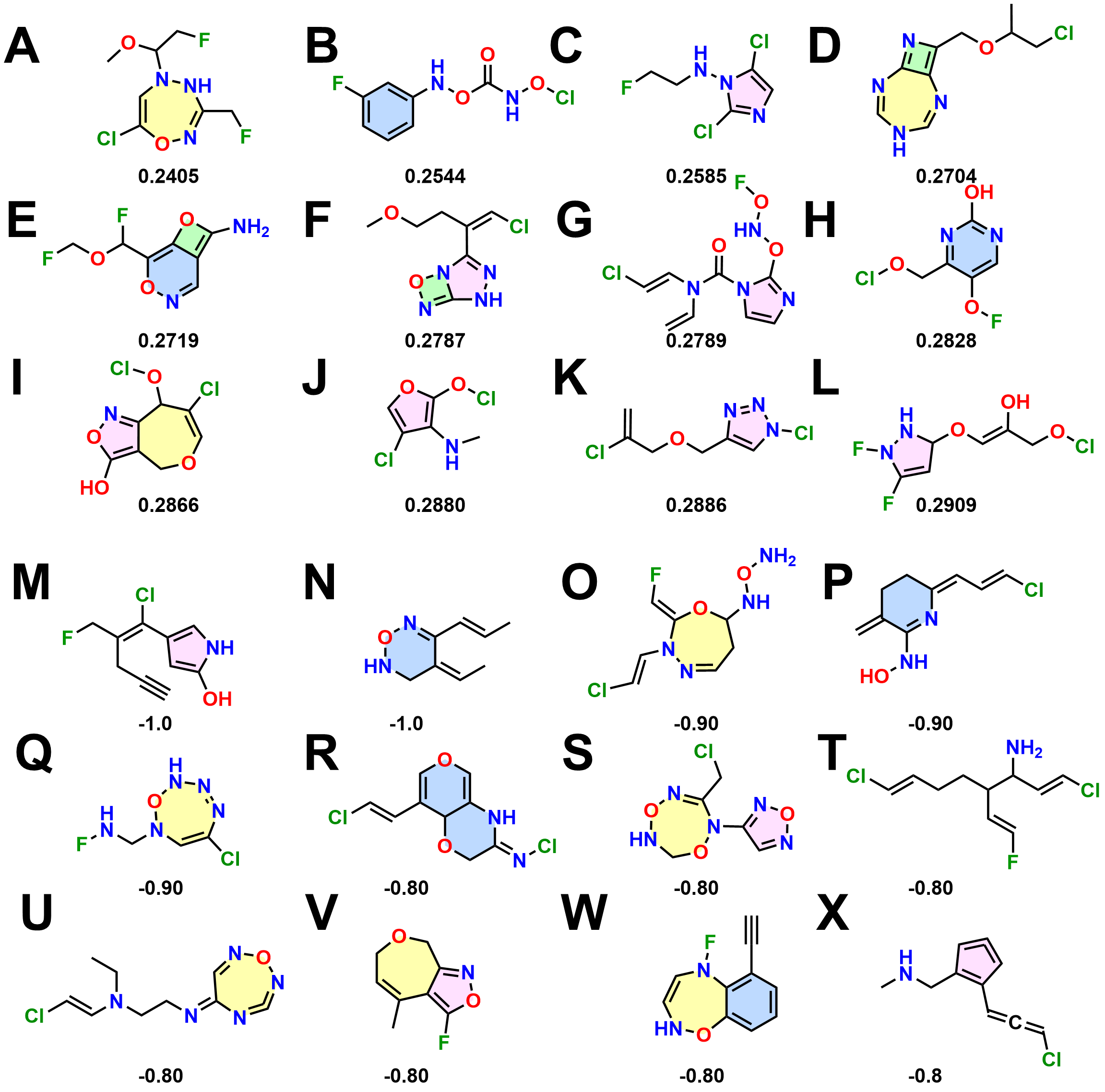}
    \caption{\textbf{Top-ranked molecules: unbiased design.} Top-12 molecules generated by QEVO as ranked according to the loss function, $\mathcal{L}$, used during the QEVO simulation (\textbf{A}-\textbf{L}) or after docking calculations (\textbf{M}-\textbf{X}). The loss value and the docking energies (in kcal/mol) are reported for each molecule.}
\label{mols_PANEL_unbiased}
\end{figure}

\begin{figure}[!htbp]
    \centering
    \includegraphics[width=1.0\linewidth]{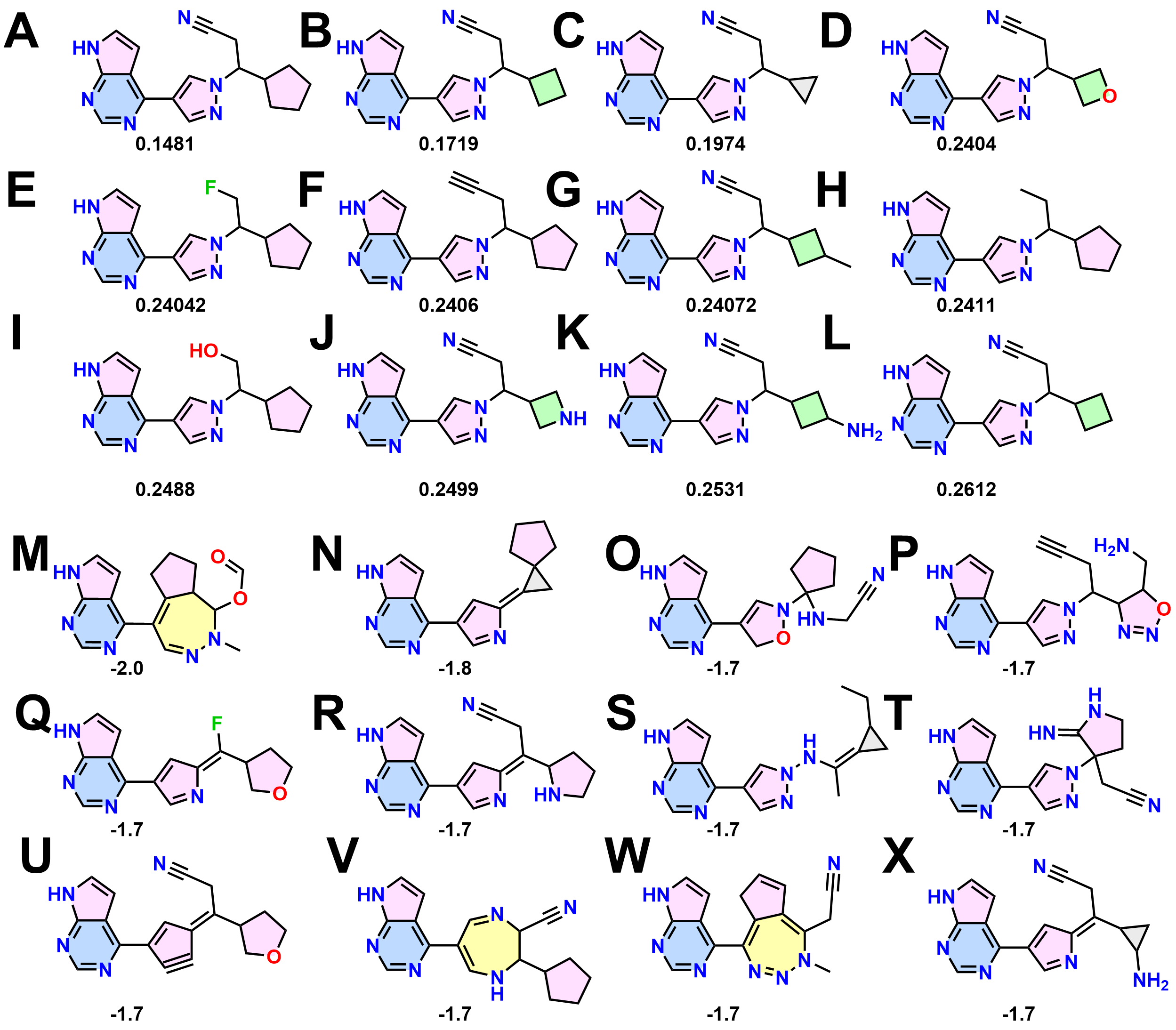}
    \caption{\textbf{Top-ranked molecules: biased design.} Top-12 molecules generated by QEVO as ranked according to the loss function, $\mathcal{L}$, used during the QEVO simulation (\textbf{A}-\textbf{L}) or after docking calculations (\textbf{M}-\textbf{X}). The loss value and the docking energies (in kcal/mol) are reported for each molecule.}
\label{mols_PANEL_biased}
\end{figure}

Figs. \ref{mols_PANEL_unbiased} and \ref{mols_PANEL_biased} show the best-scored molecules sampled by QEVO and ranked according to either their loss value or the selectivity docking score  (see Sec. \ref{methods} for computational details).
The reader will notice that not all the valid molecules reported are sound to chemists. This is not unexpected when working with unsupervised generation of SELFIES/SMILES strings. In fact, the syntactic rules internal to most popular cheminformatics libraries often ignore typically human considerations on the (geometrical) stability of molecules. 
An example of this is the SELFIES string \texttt{[C][=C][=C][Ring1][Ring1]}, {\em i.e.} \texttt{C1=C=C1} in SMILES. This string encodes a 3-carbon cycle with two C=C bonds. It is chemically meaningful since the total number of electrons and bonds per atom is correct. However, any chemist would argue about the actual stability of this molecule, and we would do as well. In the present work, such a problem is not addressed since logP, SAS, and QED are 2D properties, which means that they can be successfully computed for any valid SMILES string without affecting the validity of our results. 
Anyway, this peculiarity of SMILES encoding should be taken into consideration when 3D properties are investigated. To encourage users to verify the final chemical correctness of generated molecules, the modularity of the QEVO Python code allows the user to customize the property evaluation function of the generated SELFIES strings, thus, allowing to include any desired further validity check.

\begin{figure}[!htbp]
    \centering
    \includegraphics[width=1.0\linewidth]{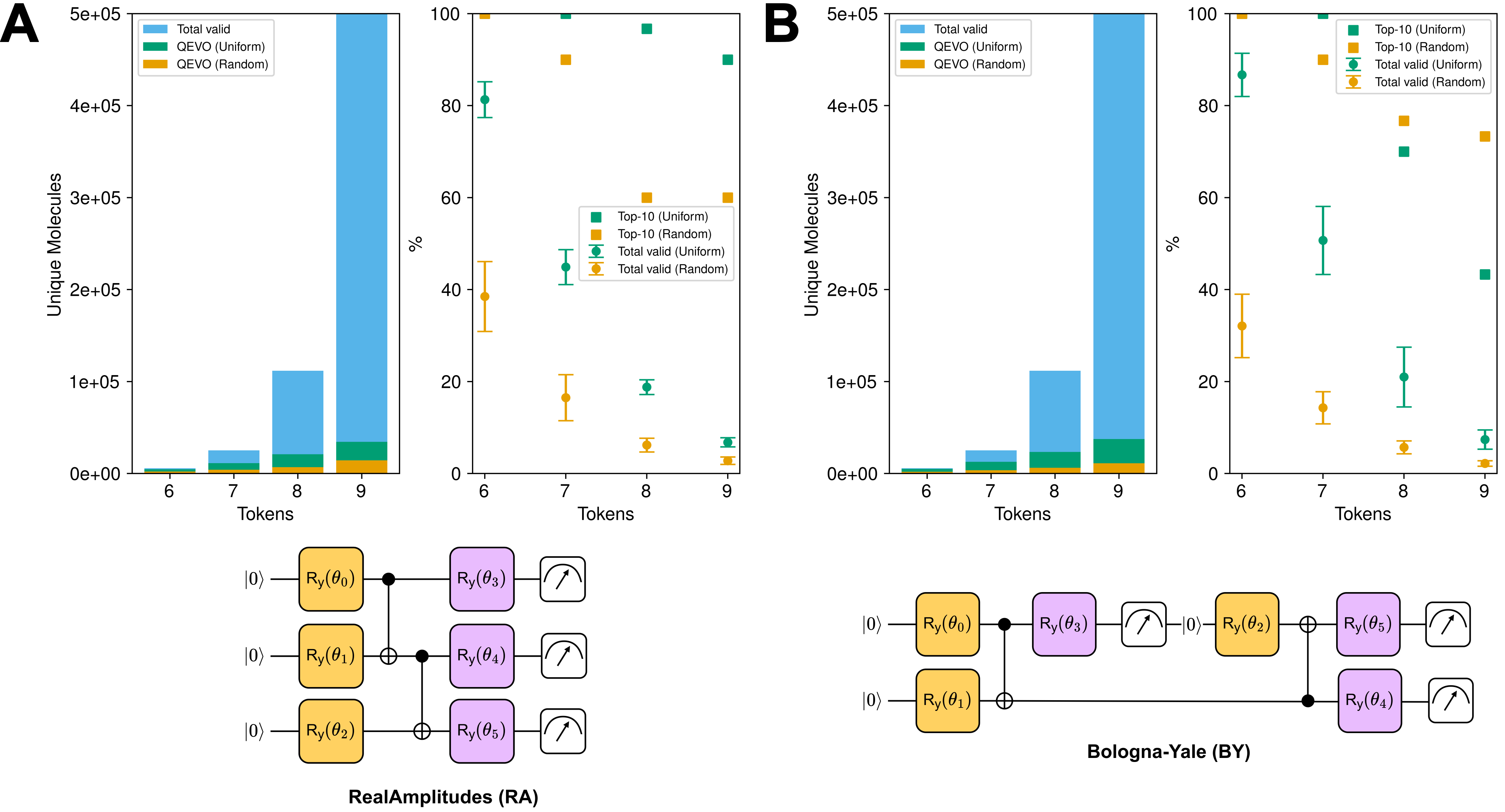}
    \caption{\textbf{QEVO performance in drug design.} The total number of valid molecules in the chemical space (blue histogram), and the number of unique molecules sampled by QEVO using either the RealAmplitudes ansatz (\textbf{A}) or the Bologna-Yale ansatz (\textbf{B}) initialized as a uniform (orange histagram), or random (green histogram) superposition are compared. The average ratio of unique molecules sampled by QEVO, after initializing the ansatz with a uniform superposition (orange round scatter), or randomly (green round scatter) scheme are reported with the success rate of generating at least one solution ranked in the top-10 solutions in the reference chemical space when either the uniform (green squared scatter) or random (orange squared scatter) is chosen. All values are averaged over 30 independent simulations and repeated for the 6-, 7-, 8-, and 9-token spaces sampling 1,024 binary strings each iteration.}
\label{trend_drug_design}
\end{figure}

\clearpage
\pagebreak

\subsection{Docking poses}

\begin{figure}[!htbp]
    \centering
    \includegraphics[width=1.0\linewidth]{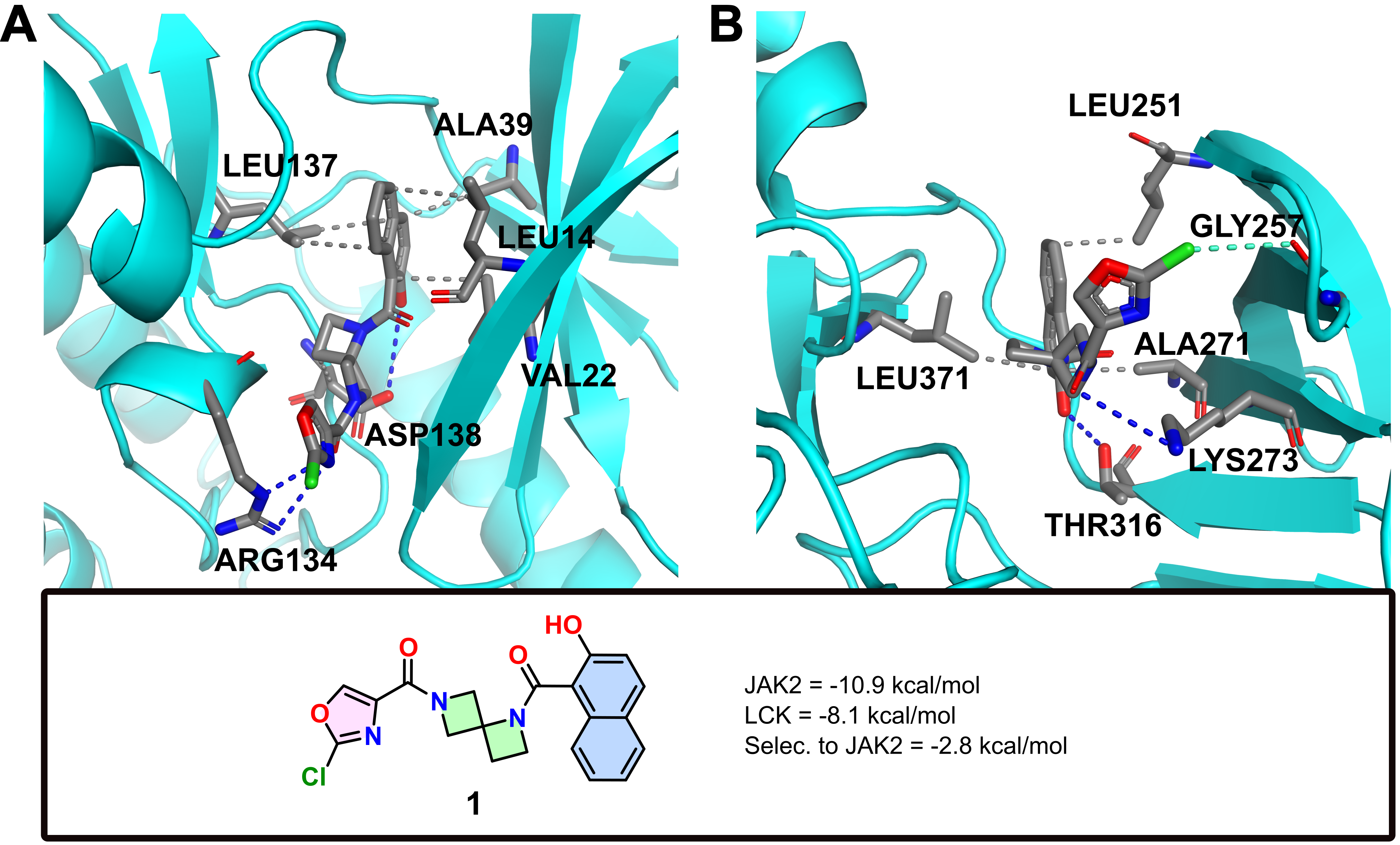}
    \caption{\textbf{Docking poses of species 1.} Docking pose analysis of the top-scoring molecule sampled from the ZINC22 database in JAK2 (\textbf{A}) and LCK (\textbf{B}). Hydrogen bond, hydrophobic interactions, and halogen bonds are displayed using blue, gray, and light green dashed lines, respectively.}
\label{docking_ZINC22}
\end{figure}

\begin{figure}[!htbp]
    \centering
    \includegraphics[width=1.0\linewidth]{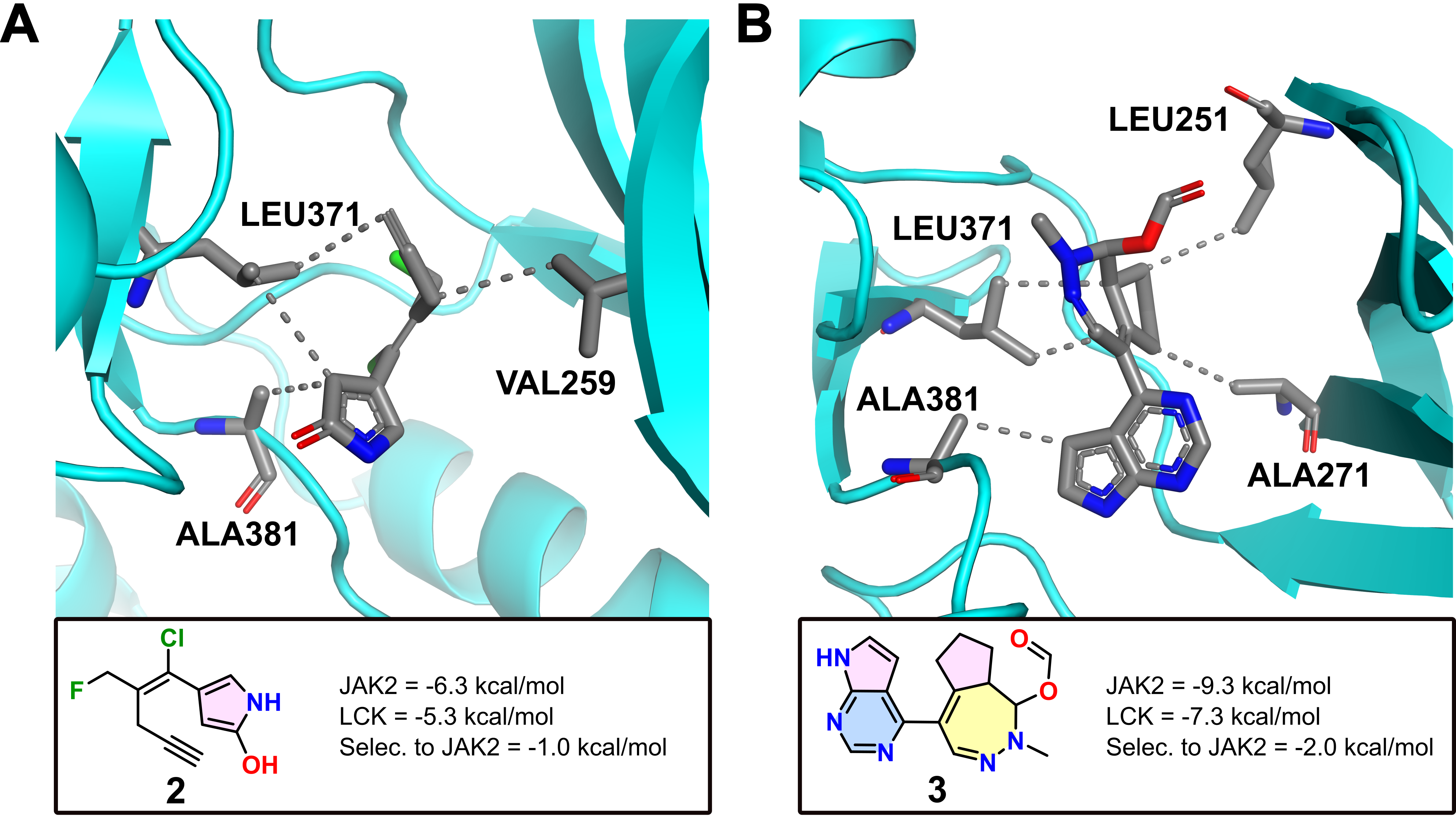}
    \caption{\textbf{Docking poses in LCK.} Docking pose analysis of the top-scoring molecules generated by QEVO both with the unbiased \textbf{A} or Tanimoto-biased \textbf{B} optimization. Hydrogen bond and hydrophobic interactions are displayed using blue and gray dashed lines, respectively.}
\label{LCK_QEVO}
\end{figure}

\pagebreak

\subsection{Convergence plots}
\begin{figure}[!htbp]
    \centering
    \includegraphics[width=1.0\linewidth]{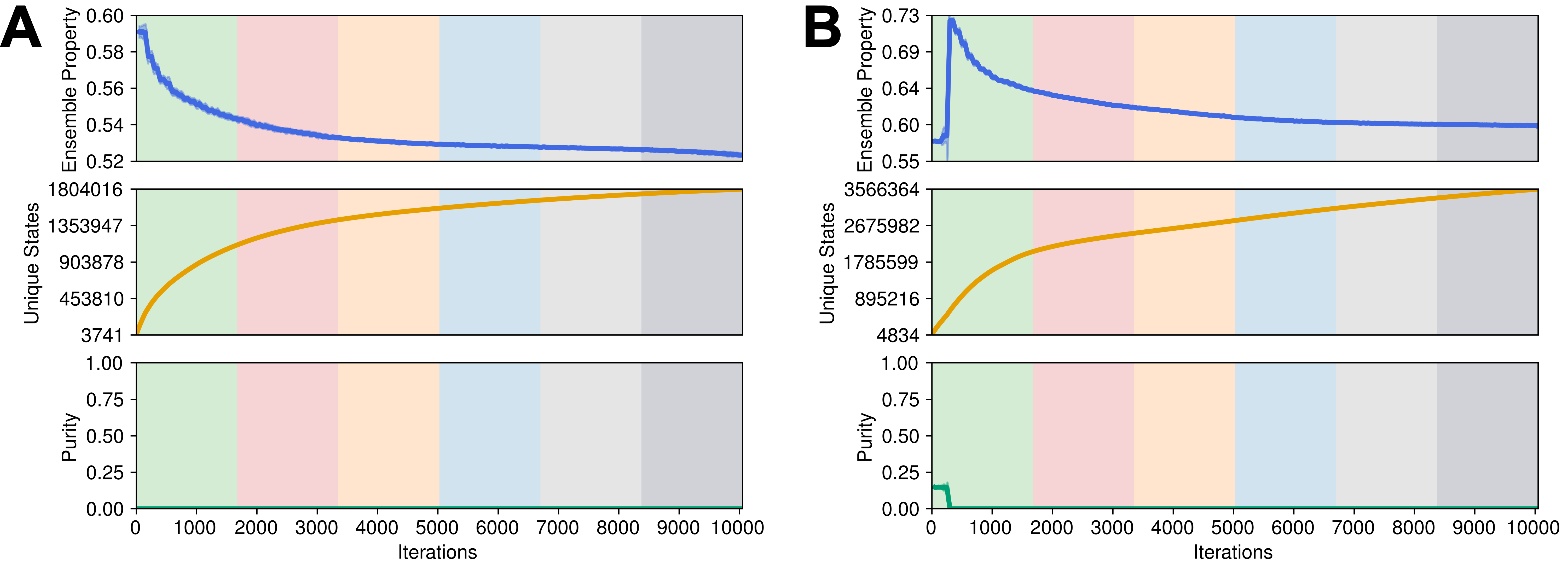}
    \caption{\textbf{Convergence plots for drug design.} The convergence of the running average (over 50 samples) ensemble property value using the QEVO algorithm (blue solid line), the cumulative number of unique combinations explored by QEVO (orange solid line), and the time-evolution of the purity of the ensemble (green solid line) are reported for the $2^{160}$ reference space using the unbiased loss function (\textbf{A}) and the one biased by the Tanimoto similarity (\textbf{B}). The total iterations are divided in 6 windows with different background colors.}
\label{drugs_designs_convergence}
\end{figure}

\end{appendices}
\newpage
\FloatBarrier
\section{References}
\renewcommand{\refname}{}
\bibliographystyle{unsrt}   
\bibliography{references}

\end{document}